
          \def\ph{\varphi}
          \def\a{\alpha}
          \def\al{\alpha}
          \def\ep{{\cal E}}
           \def\m{\medskip}

          \magnification=1200
          \bigskip
         \centerline {\bf Batalin--Vilkovisky Formalism}
                             \centerline {\bf and}
         \centerline {\bf Odd Symplectic Geometry}
                           \bigskip
                    \centerline {{\bf O.M. Khudaverdian}
                    \footnote{*} {Partially supported
                    by Grant No. M3Z300 of Inter\-na\-tio\-nal
                    Scien\-ce
                    Foun\-da\-tion}}
                           $$ $$
        \centerline{\it Department of Theoretical Physics, Yerevan State
        University}
        \centerline {A. Manoukian St.1, 375049 Yerevan, Armenia}
           \centerline {and}
           \centerline {Laboratory of Theoretical Physics,
           Yerevan Physics Institute}
           \centerline {Br. Alikhanian St.2, 375036, Yerevan
           Armenia}
           \centerline{e--mail: "khudian@sc2a.unige.ch" or
           "khudian@vx1.yerphi.am"}
                      \bigskip
            \centerline {August 1995}
                     \medskip

         $$ $$
           {\it  It is a review of some results in Odd symplectic
           geometry related to the Batalin--Vilkovisky Formalism}

      $$ $$

         \centerline {\bf Contents}
\medskip
         1. Introduction
         \medskip
         2. Batalin--Vilkovisky formalism

        $\qquad$ 2.1 Closed and open algebras of symmetries

       $\qquad$ 2.2 BV prescription

      $\qquad$ 2.3 Abelization of gauge symmetries and BV prescription
              \medskip
    3. Integration Theory over Surfaces in a Superspace

    $\qquad$  3.1 Densities in Superspace and Pseudodifferential
         forms

    $\qquad$  3.2 Dual densities and Pseudointegral forms

              \medskip
               4. Odd symplectic geometry

               $\qquad$ 4.1 Basic definitions

           $\qquad$  4.2 $\Delta$--operator in Odd symplectic geometry

           $\qquad$ 4.3 Invariant densities in Odd symplectic geometry

          $\qquad$ 4.4 $SP$ structure and BV formalism
 \medskip
           5. Acknowledgments
                  \medskip
       {\it References}
                \vfill
                \eject

               \centerline   {\bf 1 Introduction}
               \bigskip

    The last twenty years the methods dealing
    with constrained systems dynamics were essentially developed
    on the base of BRST method.  BRST method was  first introduced
    in [15,16] and [58] for treating the gauge theories and
    nowadays this method is the most powerful when dealing with
   the degenerated Lagrangians in Field Theory.

 The BRST method got very elegant mathematical formulation
  in the Hamiltonian as well as in the Lagrangian frameworks
 in the series of remarkable works of Fradkin, Batalin, Vilkovisky
 [25,26,10,27] (see also the review [31]) and [11,12,13,14] .
 ---It turns out that BRST method
 which in fact is highly developed Lagrangian multipliers
 method [42] received its mathematical formulation in terms
 of the Symplectic Geometry of Superspace.
 In general case where the algebra of symmetries of the Theory
 is not closed off--shell (i.e. the commutator of two
 infinitesimal symmetry transformations is symmetry
 transformation up to equations of motion) the
 \medskip
    Superspace $=$ Space of the initial fields $+$
       Odd Space of the ghosts fields corresponding to symmetries
       \medskip
    (Superspace= Space of the fields $+$ Odd Space
      of antifields)
    \medskip
provided with the Poisson bracket corresponding to
  Even (Odd) symplectic structure is the bag in which can be
  packed in a very compact and beautiful  way all the stuff
  (constraints, structure functions, ghosts,... ) arising
   during BRST procedure in Hamiltonian (Lagrangian)
    frameworks.
     In both approaches the application of the Symplectic
     Geometry is highly formal and technical. But there is  an
     essential difference between
      Hamiltonian and Lagrangian cases.
     One cannot say that the  necessity of
     application of Even symplectic geometry in Hamiltonian framework
     induced its development in mathematics. It is not the case
     for Odd symplectic geometry.

In the pioneer works of Batalin--Vilkovisky [11,12,13,14]
the Lagrangian  covariant formulation of BRST formalism was
constructed.
These works in  fact contain the constructions which
were the beginning of Odd symplectic
geometry.  The following mathematical constructions
  used in these works were proposed for
 mathematical investigations:

 1) The master--equation of the Theory was formulated in terms
 of Odd Poisson Bracket

 2) For formulating a Quantum Master--equation it was
 introduced the Delta--operator in the space of
 fields--antifields $(\Phi^A,\Phi^*_A)$:
                         $$
                      \Delta=
    {\partial^2\over\partial\Phi^A\partial\Phi^*_A}
                                      \eqno(1.1)
                          $$
3) It was considered the group of canonical transformations
preserving this operator---canonical transformations preserving
canonical volume form in the space of fields--antifields.
(Canonical transformations do not preserve volume form)

   \medskip

    During the  years it becomes clear that these mathematical
   constructions  are very fruitful
   for mathematical investigations.---They indeed contain
   a rich and beautiful geometry.

  This paper is mostly devoted to the geometrical problems
  arising  from the constructions of Batalin--Vilkovisky
  (BV) formalism  in the [11,12] and to the
  interpretation of the BV formalism in terms of this geometry.

  We sketch here briefly the main properties of
    Odd symplectic geometry.

 On the superspace one can consider Even or Odd symplectic
 structures given correspondingly by Even or Odd non--degenerated
 closed two form on it. The analogue of Darboux Theorem [1]
 states that there are (locally) the coordinates in which
 to Even structure corresponds Poisson bracket which conjugates
 half of bosonic
 coordinates to another half (as for usual symplectic structure
 on the underlying space) and fermionic coordinates to
 themselves. If the symplectic structure is Odd then there are
 coordinates in which Poisson bracket conjugates bosonic
 coordinates to fermionic ones (see [57]).

 There is essential difference between Even and Odd symplectic
 structures.  Even structure on a superspace
  can be considered as a natural prolongation
 of the usual symplectic structure
 from the underlying space. It is not the case for Odd one.
 Let us consider following basic example:

Let $T^*M$ be cotangent bundle of $M$ with canonically defined
symplectic structure on it [1]. By changing the parity of covectors we
come from $T^*M$ to the superspace $ST^*M$ associated with
$T^*M$. The canonical symplectic structure transforms to
Odd symplectic structure.(See for details Section 4).
The natural  correspondence between polyvectorial fields on
$M$ and the  functions on $ST^*M$ transforms Schoutten bracket
to Odd Poisson bracket\footnote{*}
{It is the reason why one of the names of Odd bracket
proposed by Leites [43,45] is Buttin bracket--- In  1969
C. Buttin in [22] investigated the graded algebras of
polyvectorial fields.}

 Indeed roughly speaking for physicists
 the supermathematics often is nothing
 but changing of small greek and latin letters on capital letters
 and putting in the suitable places the corresponding
 sign factors---powers of $(-1)$. And very often it is the fact.
 (See for example the most part of the formulae in this paper).
  But there are  cases where the constructions in
  supermathematics have the properties which radically
  differ  from the properties of their ancestors
 (in a bosonic case).
    And it is the case when we deal with Odd symplectic
    structure.

   Like for usual symplectic structure
  the group of transformations preserving Even (Odd) symplectic
  structure is infinite--dimensional: to every function
  corresponds vector field--infinitesimal transformation
  preserving symplectic structure.\footnote{**}
  { Symplectic geometry is adequate language for Hamiltonian
  Mechanics. And more natural is application of Even and Odd
  symplectic geometry for formulation of Hamiltonian mechanics in
  superspace [43]. The formulation of Hamiltonian mechanics in
  terms of even bracket describes the classical mechanics
  of fermions (See for example [18]). In the middle of
  80--th D.V.Volkov with collaborators proposed to consider
  odd symplectic structure as more fundamental for quantization.
  ([60,61], see also [36]. But till now there is
  no essential development in this direction.}
   That is why mechanics is meaningful and geometry is very poor.
    In the case of usual symplectic geometry
   canonical transformations "kill" all the invariants except
   the Liouville volume form (and corresponded Poincare--Cartan
   integral invariant). The same happens in
   supercase.

Moreover (and here begins the essential difference between Even
and Odd structures) the Odd canonical transformations on the contrary
to Even  ones do not preserve any volume form.
(If bosonic coordinate $x^1$ is multiplied by $2$ and conjugated
fermionic one $\theta^1$ is divided by $2$ then the volume form
$dx^1 d\theta^1$ is multiplied by $4$). So at first sight
the Odd symplectic structure have more poor geometry than Even
one. But the fact that no volume form is preserved by
the Odd canonical transformations makes meaningful to consider
 the superspace provided with Odd symplectic structure
 and a volume form simultaneously.
 One can consider as a group
 of transformations the group of Odd canonical transformations
 preserving this volume form. It turns out that non--trivial
 geometry is related with this structure.
 The geometrical objects depended on a higher derivatives
 appear [34,35].  Let we consider for example
 the second order operator
 with value  on a function  equal to the divergence
 (by the volume form) of the Hamiltonian vector field
 corresponding to this function via Odd symplectic structure.
 One can see that it is second--order differential
 operator which is the covariant expression of the Delta--operator (1.1)
 [34]. (The corresponding constructions
 for Even structure are trivial).
 In the special case where Delta-- operator on $ST^*M$ is
 generated by volume form on $M$ one can see that
 its action on the function  corresponds to the action of
 divergence operator on polyvector fields i.e. it is nilpotent:
                     $$
                  \Delta^2=0 .
                                        \eqno(1.2)
                     $$
         In general it is not the case.
 It turns out that

     {\it The BV master--equation can be formulated as the
     nilpotency condition of the  Delta--operator
     corresponding to the volume form (in the space of
     fields--antifields) related with the exponent
     of the master--action of the theory.}

        One has to note that in the physical examples
        of local field theories with an open algebra of the
        symmetries (such as supergravity Lagrangians)
        the Delta--operator governing
         BV-- quantum master--equation
        has a pure academical interest. The known cases
        are treated by the procedure suggested in [33,51]
        which is a special case of BV--formalism.
         During the years its geometrical properties
         were not  under the serious attention.
         Some problems of Odd symplectic geometry were
         considered in [34,35,38].

  In a [70] Witten   proposed a program for the construction of
  String Field Theory in the framework of the Batalin--Vilkovisky
  formalism and noted the necessity of its geometrical
  investigation.
    The properties of this geometry were investigated
    in [55,56], [38,39,40] and [30]. The most detailed analysis
    was performed by A.S.Schwarz [55,56].

    The BV formalism is developing now in different directions.

    The understanding of the meaning of the Delta--operator
    induces the activity for investigating the algebraical
    properties of Delta--operator and its application
    to Topological Field Theory.
    (See for example [52], [29]).
     We have to note also multilevel field--antifield
     formalism with the most general Lagrangian
     hypergauges developed by Batalin and Tyutin [7,8,9]
     and of course SP(2) BV--formalism
  (see [4,5], [6] and also [50]). It is interesting to
     note the problem of locality of the master--equation
     general solution and  the approach to the BV formalism
     based on the Koszul--Tate resolution ([42], [23], [46,21] and
     [24,32,59]). There are also
     an interesting results of application of Odd symplectic
     geometry which are not strictly connected
     with BV formalism [38,47,48,49].
      In this paper we do not consider these topics.
     Our aim is very restrictive: to give a description of the pioneer
     work of Batalin--Vilkovisky on the basis of Odd symplectic
     geometry.(We even do not consider here so called case of
     reducible theories [13], [53]).

     In the second section of this paper we give a survey
     of BV formalism making accent on its
     algebraico--geometrical meaning.

   The content of the third section is devoted to the integration
   theory over surfaces in a superspace [40].
   We consider densities---the objects which can be integrated
   over the surfaces and investigate the
    problem of defining the right generalization
   of the closed differential forms on the supercase. This
   problem indeed is strictly connected with a problem of
   reducing of partition function of degenerated theory
   on the surface of the constraints (gauge conditions).---
   From the geometrical point of view to the symmetries of a
   Theory correspond vectors fields on the space of fields which
   preserve the action. The reduced partition function,
   when gauge conditions are fixed is the integral of a
   non--local density constructed by means of these vector fields
   over the surface defined by the gauge conditions.
   The gauge independence means that this density has to be
   closed.

   In the bosonic case differential forms are simultaneously
   linear functions on the tangent vectors and well defined
   integration objects. In the supercase it is not the case.---
   The role of the differential forms as integration objects are
   played by so called pseudodifferential and pseudointegral
   forms.
    ( The investigations of these problems were started
    in a right direction in a  works
    [19,20]  then were  continued in
     [28] and [2,3] and were considered
in details  in the series of papers [62--68].)
   Our considerations in this section are based on these works.

     In the 4-th Section we deliver the main results
     in Odd symplectic geometry
     (described shortly above) related to BV formalism
  and give an interpretation of BV formalism in terms of this
  geometry.

 Our considerations  are based on the works
 [38,39,40,56] and on unpublished results of the author.

 We use the definitions and notations in supermathematics following
 [17,44,45,54]. All the derivatives in this paper are left.

\magnification=1200
\def\p{\partial}
\def\ph{\varphi}
\def\a{\alpha}
\def \al{\alpha}
\def \ep {{\cal E}}

                               $$ $$
       \centerline {\bf 2. Batalin--Vilkovisky Formalism}
\medskip
In this section we give the description of BV formalism [11,12,14]
making accents on its geometrical meaning.
          \medskip

 \centerline{2.1 Closed and open algebras of symmetries}

 \medskip
 Let ${\cal  E}$ be the space of all field configurations and a
theory be described by the action
               $$
S=S(\varphi^A),\,\varphi^A\in{\cal E}.
                                  \eqno(2.1.1)
               $$
 We use the language of de-Witt condensed notations.
 Index $A$ runs over all discrete and continuous indices
\footnote{*} {On this language the field
$\varphi(x)$ is the point $\varphi^A$ in ${\cal E}$. The
action---field dependent functional
$S=S(\varphi^A)$ is (2.1.1). The
variational derivative of the functional
${\delta S\left(\varphi^a(x)\right)\over\delta\varphi^a(y)}$
is  ${\partial S(\varphi)\over\partial \varphi^A}$.
   The expressions like $\int e^{S(\varphi^A)}D\varphi$
   (continual integral) are formal. All our considerations below
   have exact meaning in the finite--dimensional case.
   In the real (infinite--dimensional) case they need a special
   interpretation which comes from a physical context.
   The serious drawback of this language is that the difference
   between local and non--local
   functionals is not explicit in these notations.}
  ([69]).
                    $$
              {\cal F}_A=
         {\partial  S(\varphi)\over\partial  \varphi^A}=0.
                                             \eqno(2.1.2)
                   $$
are classical equations of motion which define the space $M_{st}$
of the stationary points (field configurations) of the function
$S(\varphi^A)$ (functional $S\left(\varphi^a(x)\right)$).
                        $$
            M_{st}=\left\{ \varphi^A:
            \quad {\cal F}_A (\varphi)=0\right\}.
                                               \eqno(2.1.3)
                        $$
The action $S(\varphi)$ is non--degenerated if
                $$
             corank\,
   {\partial {\cal F}_A(\varphi)\over\partial  \varphi^B}
         \vert_{ M_{st}}=0 \quad{\rm or}\quad
 Det{\partial  ^2 S\over\partial  \varphi^A \partial  \varphi^B}
\vert_{ M_{st}}\neq 0.
                               \eqno(2.1.4)
                 $$

In a general case if (2.1.4) does not hold
 the theory is degenerated.

Let ${R^A_\alpha }$ be a set of vector fields---
 symmetries of the
 theory
                        $$
               R_\alpha ^A {\cal F}_A=0
                                                \eqno(2.1.5)
                        $$
                        $$
       {\rm i.e.}\,
       S(\varphi^A+\delta\varphi^A)-S(\varphi^A)\approx 0
              \quad{\rm for}\,{\rm infinitesimal}
                  \,{\rm variations}\quad
      \delta\varphi^A=\delta\xi^\alpha  R_\alpha ^A
                                                \eqno(2.1.6)
                        $$
which do not vanish "classically"
                        $$
     R^A_\alpha |_{M_{st}}\neq 0\,.
                                  \eqno(2.1.7)
                        $$
(2.1.5) are Noether identities of second kind.
 ($S(\varphi)$) is local functional:
                        $$
                   S(\varphi^A)=
                   \int{\cal L}(\varphi^a(x),
     {\partial\varphi^a(x)\over\partial x^\mu},\cdots)d^4x \quad
    (A=(a,x^\mu))
                                                \eqno(2.1.8)
                                $$
The global symmetries (when $\delta\xi$ in (2.1.6) do not depend
on $x^\mu$) do not put identities (2.1.5) on the motion equations
(2.1.2) (See in details [69])

The global symmetries are excluded out of consideration. If
 theory is not degenerated then (2.1.4) leads to
                        $$
              {\rm dim}\,M_{st}=0
                                              \eqno(2.1.9)
                       $$
for (2.1.3).

Of course (2.1.9) follows from (2.1.4) only if we consider the
solutions of (2.1.2) obeying to the initial conditions which
exclude the global symmetries. It is the  case when we consider a
continual integral
                        $$
Z(J)=\int e^{{1\over\hbar} S(\varphi)+
 J\varphi} {\cal D}\varphi
                                              \eqno(2.1.10)
                        $$
which yields the Green functions of the theory.

In the case if (2.1.4) ((2.1.9)) obeys, (2.1.10) can be calculated
perturbatively in power series on $\hbar$ by
 extracting the quadratic
part of the action and calculating the corresponding Gaussian
integral---It corresponds to the expansion of the action
$S(\varphi)$ around the set of stationary points---$M_{st}$.

 It is easy to see that the vector fields:
                        $$
           R^A=E^{AB}{\cal F}_B
                                           \eqno(2.1.12)
                        $$
where $E^{AB}$ is arbitrary antisymmetric tensor
                        $$
                    E^{AB}=-E^{BA}
                                        \eqno(2.1.13)
                         $$
evidently obeys to (2.1.5) and do not obeys to
(2.1.7)---it is the
symmetries vanishing on classical level.
  \footnote{*} { We often omit the sign factor in
the  formulae---i.e. the corresponding expressions are exact
 in the case where the space ${\cal E}$ of the fields is bosonic.
  For example in (2.1.13) one have to add the sign
factor $(-1)^{p(A)p(B)}$}

   One can see that if two vector fields--symmetries $T^A$
and $T^{\prime A}$ obey to (2.1.5) and coincide on $M_{st}$
                        $$
          T^A{\cal F}_A=
       T^{\prime A}{\cal F}_A=0,
                                      \eqno(2.1.14)
                         $$
                         $$
         T^A\approx T^{\prime A}\quad{\rm i.e.}\quad
           T^A\vert_{M_{st}}=T^{\prime A}\vert_{M_{st}}
                                        \eqno(2.1.15)
                          $$
then there exist $E^{AB}$ obeying (2.1.13) such that
                          $$
          T^A-T^{A^\prime}=E^{AB}{\cal F}_B\,.
                                        \eqno(2.1.16)
                          $$
We consider so called irreducible theories
and assume that the set $\{{\bf R}_\alpha\}$
of the symmetries is complete:
                   $$
             \sum_\alpha
\lambda^\alpha  {\bf R}_\alpha \approx 0 \,\Rightarrow\,
\forall \alpha \,\,\lambda^\alpha \approx 0
                                        \eqno(2.1.17)
                   $$
and
                   $$
      \forall T^A:\quad
                  \sum_A
         T^A{\cal F}_A\approx 0\quad
                \Rightarrow\,
       T^A\approx \sum_\alpha\lambda^\alpha  R_\alpha ^A .
                                \eqno(2.1.18)
                   $$
The set $\{{ R}_\alpha^A\}$ obeying to the conditions
(2.1.17) and
(2.1.18) we call the basis of the symmetries of the theory.

{\it {The "number" of symmetries of irreducible theory is equal to
dimension of $M_{st}$}}
 \smallskip
It is useful to represent the considerations above in the
following exact sequence:
                        $$
     0\rightarrow F\rightarrow E \rightarrow
                  B\rightarrow0
                                        \eqno(2.1.19)
                         $$
where $F$ is the space of symmetries vanishing on the $M_{st}$
(2.1.12) ("on--shell"vanishing symmetries), $E$ is the space of
all vector fields obeying to (2.1.5) (symmetries) and
                           $$
                         B=E/F
                                        \eqno(2.1.20)
                           $$

 {\it $B$ is the space of the symmetries of classical theory.}

$E$ and $F$ are the moduli on the algebra of the functions on
${\cal E}$. We have to note that the sequence
$$0\rightarrow F\rightarrow E \rightarrow
B\rightarrow 0$$
is typical for the theories of constrained systems. The fact that
 the
"physical space" is $B$ and on other hand the space $E$ is
preferable to work in, is the source of arising the ghosts in the
formalisms of these theories (see [42,23,24,32]).

The set of equivalence classes $\{[{\bf R}_\alpha ]\}$
consist the basis in $B$
and $\{{\bf R}_\alpha \}$  are the representatives of this basis in $E$.
(The basis of symmetries $\{{\bf R}_\alpha \}$)
defined above is the set
of representatives of the basis $\{[{\bf R}_\alpha] \}$ in $B$.

 It is easy to see using (2.1.5) that
 commutator of two symmetries ${\bf R}_\alpha$ ,
${\bf R}_\beta$  $[{\bf R}_\alpha ,{\bf R}_\beta]$ is the
symmetry too. So comparing (2.1.5), (2.1.12) and (2.1.16)
 we see that
                        $$
           \left[{\bf R}_\alpha,  {\bf R}_\beta \right]=
           t^\gamma_{\alpha  \beta}R_\gamma+
             E_{\alpha  \beta}^{AB}{\cal F}_B
                                  \eqno(2.1.21)
                        $$
Where $E_{\alpha  \beta}^{AB}$ are obeyed to (2.1.13). In the case if
                        $$
             E_{\alpha  \beta}^{AB}=0
                                     \eqno(2.1.22)
                         $$
the algebra of symmetries of the theory in physics is called
"closed algebra"("off--shell algebra of symmetries"). In the
case if (2.1.22) does not hold the algebra of symmetries of theory
is called "open algebra" ("on--shell algebra of symmetries").

Of course these definitions are ${{\bf R}_\alpha }$--basis dependence.
The space $B$ defined by (2.1.20) is in usual sense the
algebra Lie, because $F$ is ideal in $E$ as algebra of vector
fields.  It is easy to see that the transformation
                     $$
    {\bf R}_\alpha\rightarrow
  \lambda_\alpha^\beta{\bf R}_\beta+
             E_\alpha^{AB}{\cal F}_B
                                                \eqno (2.1.23)
                     $$
 where $E_\alpha^{AB}$ is antisymmetric
 (See (2.1.13)) changes the basis of symmetries
to another one. In principal by this transformation one can
construct
the basis of symmetries for which $E^{AB}_{\alpha\beta}$ in
(2.1.21) and even
$t_{\alpha  \beta}^\gamma$ is vanished---
so called {\it abelian basis
of symmetries} (See  subsection 2.3).

But in field theory we are restricted in a choosing arbitrary
basis ${{\bf R}_\alpha }$ of symmetries
(the representatives ${{\bf R}_\alpha }$)
for the basis ${[{\bf R}_\alpha ]}$) in $B$.
These restrictions are locality conditions on  ${\bf R}_\alpha $.

           $$ $$

\centerline         {2.2 BV prescription}
\medskip
For calculating the (2.1.10)---the generating functional for Green
functions in the case if theory is degenerated
   ($dim M\vert_{st}\neq 0$) one have to
exclude the degrees of freedom connected with the  symmetries
(2.1.5), (2.1.7).

If the basis of symmetries ${{\bf R}_\alpha }$ is local and abelian
 the gauge degrees of freedom are easily
extracted from (2.1.10). If the basis of symmetries
$\{{\bf R}_\alpha\}$
consist the Lie algebra ($t_{\alpha \beta}^\gamma\equiv
const,\,E=0$ in (2.1.21)) then we come to well--known
Faddeev--Popov  trick.

The BV--prescription for calculating the generating functional
(2.1.10) works in a most general case (2.1.21). We recall here
briefly this prescription and give in the next subsection the
arguments explaining it.

For the degenerated theory with action $S(\varphi)$ and with
basis of symmetries $\{{\bf R}_\alpha \}$ let equations
                           $$
                    \Psi^\alpha =0
                                                \eqno (2.2.1)
                            $$
   define the surface $\Omega$
 in the space ${\cal E}$ of fields
  which defines gauge conditions corresponding to the
 symmetries $\{{\bf R}_\alpha \}$
  ($dim (M\vert_{st}\cap\Omega)=0$). To reduce the continual integral
                        $$
         Z=\int e^{S(\varphi)\over \hbar}{\cal D} \varphi
             \quad ({\cal D}\ph=\prod_A d\varphi^A)
                                        \eqno(2.2.2)
                        $$
to the integral defined on this surface (the eliminating the gauge
degrees of freedom) one have consider the following
 construction [11]:

 Let ${\cal E}^e$ be a space with coordinates
                        $$
 \Phi^A=(\varphi^A, c^\alpha ,\nu_\beta, \lambda_\sigma )
                                        \eqno(2.2.3)
                        $$
where auxiliary coordinates $c^\alpha $, $\nu_\beta$
 are ghosts corresponding to the symmetries
 ${\bf R}_\alpha$,  $\lambda_\alpha $--Lagrange multipliers
 corresponding to
constraints (gauge conditions) $\Psi^\alpha $.
  The parity of Lagrange multipliers coincide and
the parity of ghosts is opposite to the parity of
corresponding symmetry:
                            $$
               p(c^\alpha )=p(\nu_\alpha)=
          p(\lambda_\alpha )+1=p({\bf R}_\alpha)+1.
                                                \eqno(2.2.4)
                             $$
 We introduce a space of fields and antifields
  $S{\cal E}^e$ with coordinates
$(\Phi^A,\Phi^*_A)$ where $\Phi^*$ have opposite parity
 to $\Phi$:
                          $$
               p(\Phi^*_A)=p(\Phi^A)+1.
                                               \eqno(2.2.4a)
                           $$
   It is convenient to consider the subspace ${\cal E}^e_{min}$
 of ${\cal E}^e$ containing the fields
 $\Phi^A=(\varphi^A,c^\alpha)$ and correspondingly
  a subspace $S{\cal E}^e_{min}$ of $S\ep^e$--- the space of
 $(\Phi^A_{min},\Phi^*_{Amin})-(\ph^A,c^\alpha,\ph^*_A,c^*_\alpha)$.

In the space of fields antifields  one have to define
the odd symplectic structure (see
for details the Section 4) by Poisson bracket
                        $$
     \{F,G\}={\partial F\over \partial \Phi^A}
     {\partial G\over \partial \Phi^*_A}
                    +
                (-1)^{p(F)}
     {\partial F\over \partial \Phi^*_A}
     {\partial G\over \partial \Phi^A}
                                        \eqno(2.2.5)
                        $$
and Delta--operator\footnote{*}{all the derivatives are left}
                        $$
\Delta F={\partial^2 F\over\partial \Phi^A\partial \Phi^*_A}
                                \eqno(2.2.6)
                        $$
Then one have to define the master--action----the function
             ${\cal S}(\Phi,\Phi^*)$
 obeying to equation
                    $$
           \Delta e^{{\cal S}\over\hbar}=0
           \Leftrightarrow
         \hbar\Delta {\cal S}+
     {1\over2}\{{\cal S},{\cal S}\}=0
                                        \eqno(2.2.7)
                     $$
       or classically
                      $$
               \{{\cal S},{\cal S}\}=0
                                         \eqno(2.2.7a)
                      $$

 (the term proportional to $\hbar$ in (2.2.7) is responsible to
 measure factor.)

and to initial conditions which are defined by the action $S(\ph)$
and symmetries ${\bf R}_\al$:
                      $$
            {\cal S}\vert_{\Phi^*=0}=S(\varphi),\quad
                  {\partial^2 {\cal S}
                     \over
          \partial c^\alpha\partial\varphi^*_A }
                     \vert_{\Phi^*_A=0}=R^A_\alpha,
                     \,\,
                   {\cal S}(\Phi,\Phi^*)=
    \nu^{*\beta}\lambda_\beta+{\cal S}(\Phi_{min},\Phi^*_{min})
                                                       \eqno(2.2.8)
                      $$
  i.e.
                     $$
                   {\cal S}=
              {\cal S}(\ph,c,\ph^*,c^*)+
            \nu^{*\alpha}\lambda_\alpha=
    S(\varphi^A)+c^\alpha  R^A_\alpha \varphi_A+
     \dots+\nu^{*\alpha}\lambda_\alpha\,.
                                        \eqno (2.2.8a)
                         $$
 (The dependence of ${\cal S}(\Phi,\Phi^*)$ on the fields
 $(\lambda,\nu,\lambda^*,\nu^*)$ is trivial)
  The equation (2.2.7) is called "master--equation".
 It can be proved that the master--equation with boundary conditions
 (2.2.8) have unique solution [14].

To gauge fixing conditions corresponds gauge fermion
                        $$
           \Psi=\Psi^\alpha  \nu_\alpha
                                        \eqno(2.2.9)
                        $$
The partition function (2.2.2) is reduced to integral
                        $$
                    Z^{\prime}=
                       \int
              e^{{\cal S}(\Phi.\Phi^*)}
                  \delta\left(
     \Phi^*_A-{\partial\Psi\over\partial\Phi^A}
              \right)\prod_A d\Phi^A d\Phi^*_A
                                                \eqno(2.2.10)
                          $$
To the changing of gauge (2.2.1) corresponds the changing of $\Psi$
in (2.2.9). The integral (2.2.10) does not depend on the
choice of $\Psi$.
(Later we will discuss the geometrical meaning of this construction).

In the case if basis of symmetries ${{\bf R}_\alpha }$ consists Lie
algebra one can show that
                $$
                {\cal S}=
               S(\varphi)+
               c^\alpha R^A_\alpha \varphi^*_A+
                 {1\over2}
            t^\alpha_{\beta\gamma}c^*_\alpha  c^\beta c^\gamma
                  +\nu^{*\alpha}\lambda_\alpha
                                                    \eqno(2.2.11)
                     $$
and (2.2.10) reduces to well--known Faddeev--Popov trick.

In the next section we deliver arguments explaining these
constructions.

 \bigskip

  \centerline{ 2.3 Abelization of Gauge Symmetries and BV
  prescription}
  \smallskip
  $\qquad$ $\qquad$  $\qquad$  $\qquad$  $\qquad$
  $\qquad$ $\qquad$ $\qquad$ {\it "Make straight the way of the Lord"}

   $\qquad$ $\qquad$  $\qquad$ $\qquad$ $\qquad$
   $\qquad$ $\qquad$  $\qquad$ $\qquad$ $\qquad$
                        ( St John {\bf 1}: 23)
     \medskip
    In this subsection we will give motivation for BV
    prescription
    and will see how the odd symplectic structure arise
    in this procedure. Our considerations in this
    subsection are based on [12].
    In 4--th Section we will
    study this problem on the background of odd symplectic
    geometry.

 Let us consider first a simplest case where $\{{\bf R}_\alpha\}$ is
abelian basis of symmetries.
                                $$
           \left[{\bf R}_\alpha,{\bf R}_\beta\right]=0.
                                                \eqno(2.3.1)
                                  $$
 We will show below that in this case
  the eliminating of gauge
degrees of freedom reduces the partition function (2.2.2) to the
                     $$
          Z^{\prime}=\int
      e^{S(\varphi)}
          {\rm Det}\left(R_\alpha^A\,
{\partial \Psi^\beta\over\partial  \varphi^A}\right)
        \prod_\alpha\delta(\Psi^\alpha)
              \prod_A d\varphi^A
                                        \eqno(2.3.2)
                          $$


Indeed even in the case where basis of symmetries forms  Lie
algebra, (2.3.2) gives correct answer for the partition function.
The localizing of nonlocal functional
 ${\rm Det}\left(R_\alpha^A
{\partial \Psi^\beta\over\partial  \varphi^A}\right)$ in the enlarged space
of ghosts
                       $$
               {\rm Det}\,
                \left(
                R_\alpha^A\,
         {\partial \Psi^\beta\over\partial  \varphi^A}
                \right)=
                \int e^
              {c^\alpha R_\alpha^A\,
          {\partial \Psi^\beta\over\partial  \varphi^A}
                \nu_\beta}
                 \prod_\alpha dc^\alpha d\nu_\alpha
                                        \eqno(2.3.3)
                         $$
gives us well--known Faddeev--Popov trick.

  (The geometrical meaning of (2.3.2) and of (2.3.3) see in 3-th
  Section)

 Before going in
delivering the eq. (2.3.2) we will show that
 it coincides with BV partition function
(2.2.10).

 Indeed in the case (2.3.1) the solution of (2.2.7) is
                        $$
{\cal S}=S(\varphi)+c^\alpha R^A_\alpha\varphi^*_A+
  \nu^{*\alpha}\lambda_\alpha
                                        \eqno(2.3.4)
                        $$

 Indeed it is easy to see that in this case

			$$
         \{{\cal S},{\cal S}\}=
        2R_\alpha^A{\p S\over \p \ph^A}=0.
				\eqno(2.3.5)
			$$

(We consider the case where
			$$
         {\p R^A\over \p \ph^A} = 0
				\eqno(2.3.6)
			$$
(the symmetries preserve volume form). See also remark after
 (2.3.15)).

In this case using (2.3.3) we can rewrite (2.3.2) in the form (2.2.10)
			$$
              \int e^{S(\ph)}
          \prod_\alpha\delta(\Psi^\alpha)
         Det(R^A_\alpha{\p\Psi^\beta\over\p\ph^A})
                      \prod_A d\ph^A=
			$$
			$$
                   \int e^
                        {
               S(\ph)+\lambda_\alpha\Psi^\alpha_+
        c^\alpha R^A_\alpha{\p\Psi^\beta\over\p\ph^A}\nu_\beta
                        }
                \prod_{A,\alpha}
           d\lambda_\alpha d\nu_\alpha dc^\alpha d\ph^A=
			$$
			$$
                  \int e^
                     {
         S(\ph)+c^\alpha R^A_\alpha\ph^*_A+
           \nu^{*\alpha}\lambda_\alpha}
                \delta(\ph_A^*-{\p\Psi\over\p\ph^A})
            \delta(\nu^{\a*}-{\p\Psi\over\p\nu_\a})
                   \delta(c_\a^*)
                 \prod_{A,\a} d\lambda_\alpha
                     dc^\a dc^*_\a
                  d\nu_\a d\nu^{\a*}
                   d\ph^A d\ph^*_A=
			$$
			$$
           \int e^{{\cal S}(\Phi,\Phi^*)}
       \delta(\Phi_A^*-{\p \Psi\over\p\Phi^A})
                \prod_A d\Phi^A d\Phi^*_A
                                                      \eqno(2.3.7)
			$$
where $\Phi^A=(\ph^a,c^\a,\nu_\a,\lambda_\a)$ is given by
 (2.2.3) and $\Psi$ is given by (2.2.9).

  In general case (2.1.21), (2.3.2) depends on the gauge
  conditions (2.2.1) because the integrand in (2.3.2) is not anymore
  closed density (see Section 4).
  For obtaining (2.2.10) we do following:

1) From the basis of symmetries
${\bf \{R_\a\}}$
we go to abelian basis of symmetries
${\bf{\cal \{R_\a\}}}$
(temporary ignoring the problem of locality of symmetries)

2) We will show that in abelian basis we will come to (2.3.2) -
so (2.2.10) is valid in this case (See eq.(2.3.7) above).

3) Then we will return from non-local abelian
basis ${\bf{\cal \{R_\a\}}}$
to local physical basis ${\bf \{R_\a\}}$.
We will see that in the enlarged
space $S{\cal E}^e$ of the fields-antifields the returning to initial
symmetries corresponds   to the canonical transformation
preserving (2.2.5)
and master-equation (2.2.7). Using uniqueness of the solution of (2.2.7)
with boundary condition (2.2.8) we come to (2.2.10).
 \medskip
1) Let ${\bf \{R_\a\}}$ be basis of symmetries of theory $S(\ph)$. Let
$\xi^a$ be the coordinates on some surface $\Omega_0$ given by the equation
			$$
                     \Psi^\a_0=0
							\eqno(2.3.8)
			$$
which is transversal to vector fields ${\bf \{R_\a\}}$.
 One can introduce
 in the space ${\cal E}$ the new coordinates
 $(\xi^a,\eta^\a)$, which
 correspond to symmetries ${\bf \{R_\a\}}$:
for every set $(\xi^a_0,\eta^\a_0)$ we consider the
integral curve (the exponent)
of vector field ${\bf R}(\eta_0)=\eta_0^\a {\bf R}_\a$:
			$$
                       \eqalign
                         {
                 \gamma_{\eta^\a_0}(t)& =
               \exp(t\eta^\a_0 {\bf R}_\a)
                  |\ph_0\rangle, \qquad (0\leq t\leq 1)
                          \cr
{d\gamma_{\eta^\a_0}(t)\over dt}&= \eta^\a_0 R_\a (\gamma)
                           }
						\eqno(2.3.9)
			$$
beginning at the point $\ph_0$ with coordinates
 $\xi^a_0$ on the surface $\Omega_0$. To the ending point
of this curve corresponds the set
$(\xi^a_0,\eta^\a_0)$.

Of course, these new coordinates in general are non-local. But we do not
pay attention on this fact because at very end we return to initial local
coordinates.

It is evident that the action $S$ does not change along the integral curves
$\gamma^A(t,\xi,\eta)$ so in the new coordinates, $S$ does not depend
on $\eta^\a$
			$$
                     S=S(\xi^\a)
						\eqno(2.3.10)
			$$
and ${\bf {\cal R_\a }}=\{{\p\over \p\eta^\a}\} $ is evidently the abelian
basis of symmetries. In the initial coordinates $\ph^A$
this abelian basis is equal to
			$$
{\bf{\cal R}_\a}={\p\over \p\eta^\a}= {\p\ph^A(\xi,\eta)\over  \p \eta^\a }
{\p\over\p\ph^A},
						\eqno(2.3.11)
			$$
			$$
{\cal R_\a^A}={\p\ph^A\over \p\eta^\a}.
					       \eqno(2.3.12)
			$$

In the coordinates $(\xi^a,\eta^\a)$ the problem of excluding the gauge
degrees of freedom is trivial:
			$$
                   Z^{\prime}=\
                 \int e^{S(\xi)}\prod_a d\xi^a=
               \int e^{S(\xi)}
              \prod_{a,\a}\delta(\Psi^\alpha)
                     \vert
           {\partial\Psi^\alpha\over\partial\eta^\beta}
                      \vert
                      d\xi^a d\eta^\a .
						\eqno(2.3.13)
			$$
  Using that $R_\a^\beta=\delta_\a^\beta$, $R_\a^a=0$
 in these coordinates we come to (2.3.2).
 Master--action in these coordinates is
                         $$
{\cal S}=S(\xi)+c^\a \eta_\a^* +\nu^{*\a}\lambda_\a\,.
			$$
  In the initial coordinates $(\ph^A)$
			$$
                     \eqalign
{
                       Z^{\prime}=&
           \int e^{S(\xi)}
              \prod_\a\delta(\Psi^\alpha)
                     \vert
           {\partial\Psi^\alpha\over\partial\eta^\beta}
                      \vert
             \prod_{a,\a} d\xi^a d\eta^\a=
                       \cr
&               \int e^{S(\xi)}
              \prod_\a\delta(\Psi^\alpha)
                     \vert
           {\partial\Psi^\alpha\over\partial\eta^\beta}
                      \vert
             \prod_{A} d\ph^A
                           }
						\eqno(2.3.14)
			$$

Using (2.3.12) we come to (2.3.2):
                     $$
          Z^{\prime}=\int
               e^{S(\varphi)}
               {\rm Det}
              \left
               ({\cal R}^A_\a
            {\partial \Psi^\a\over\partial  \ph^A}
                \right)
                \prod_\alpha \delta (\Psi^\alpha)
                  \prod_A d\varphi^A
                                        \eqno(2.3.15)
                          $$
We see (using (2.3.7)) that in the basis
$\{{\bf{\cal R}}_\a\}$
(2.2.10) is valid.

The basis is abelian, exponent of action evidently obeys to master-equation.
But the price for receiving this simple formula is very high: the symmetries
${\cal R^A_\alpha}$ are nonlocal.

{\bf Remark.} Our considerations in this section are precise up to the changing
of volume form. It corresponds to the classical limit ($\hbar \to 0$)
of master equation (2.2.7a).
\medskip
3) The returning to initial symmetries $\{{\bf{\cal R}}_\a\}$:
 It is here where canonical structure plays crucial role:
The relation between new abelian basis  $\{{\bf{\cal R}}_\a\}$
and initial one is given by equation
			$$
{\bf R}_\a=\lambda_\a^\beta {\bf{\cal R}} _\beta +
     E^{{[}AB{]}}_\alpha {\cal F}_B
				\eqno(2.3.16)
			$$
(See equation (2.1.17, 2.1.18)).

One can show that the transformation (2.3.16) can be realized
 {\it by canonical
transformation in the space of fields, antifields.}
We will show it infinitesimally. We note (see in details section 4 )
that to arbitrary odd function
                  $$
           Q(\Phi_{min},\Phi_{min}^*)=
           Q(\ph,c,\ph^*,c^*)
                $$
 corresponds canonical infinitesimal transformation:
			$$
\eqalign{
	\delta\Phi^A=\epsilon\{Q,\Phi^A\} \cr
        \delta\Phi^*_A=\epsilon\{Q,\Phi^*_A\}
                    }		     \eqno(2.3.17)
			$$
  and:
 			$$
                  \delta{\cal S}=
 	         \epsilon\{Q,{\cal S}\}\,.
                                           \eqno(2.3.18)
                          $$
 If we consider
                         $$
Q=c^\alpha\lambda^\beta_\alpha c^*_\beta+
 c^\alpha c^\beta E^{AB}_{\alpha\beta}\ph^*_A\ph^*_B
                          $$

 then putting (2.3.4) in (2.3.18) and using (2.3.17) we
  obtain that
                      $$
    {\cal S}\rightarrow  {\cal S}+\delta {\cal S}=
   S(\varphi)+c^\alpha
               \left(
         {\cal R}^A_\alpha\varphi^*_A+
                \epsilon\lambda_\a^\beta{\cal R}_\beta^A+
           \epsilon E_\a^{AB}{\cal F}_B
                       \right)
                           +\dots
                                       \eqno(2.3.19)
                              $$
   Using (2.2.8a) we see that (2.3.19) corresponds to infinitesimal
transformation (2.3.16).

  We note that if the generator $Q$ of canonical transformation
obeys to equation
                           $$
                         \Delta Q=0
                                                   \eqno(2.3.20)
                           $$
then one can see  that the canonical transformation (2.3.17)
preserves volume form $dv=\prod_A d\Phi^A d\Phi^*_A$.
 Indeed from (2.3.17) follows that
                      $$
               \delta dv=0\quad
                        {\rm if}\quad
                   \Delta Q=0.
                                                    \eqno(2.3.21)
                      $$
  The classical master--equation (2.2.7a) is
invariant under canonical transformations
(transformations preserving
 odd bracket $\{\,,\,\}$), the  quantum master--equation
 (2.2.7) is invariant under the canonical transformations
 preserving the volume form.
  So from the fact that to the changing of the basis
of the symmetries corresponds canonical transformation
(canonical transformation preserving the volume form)
and from the fact that master--equation
have unique solution follows (2.2.10).

\magnification=1200


                        $$ $$
  \centerline{\bf 3.Integration Theory over Surfaces in the Superspace}
                            \medskip
   \centerline {3.1.Densities in the superspace and
   Pseudodifferential Forms}
   \medskip
In this section we present some results of geometric integration
theory on the surfaces in the superspace (see [28], [64], [40]).

Let $\Omega^{m.n}$ be an $(m.n)$--dimensional supersurface in the
superspace $E^{M.N}$ given by parametrization
 $z^A=z^A(\zeta^B)$ the mapping of superspace $E^{m.n}$ in
superspace $E^{M.N}$ where $z^A=(x^1,\dots,x^M,\theta^1,\dots,
\theta^N)$
are coordinates of superspace $E^{M.N}$ and
$\zeta^B=(\xi^1, \dots, \xi^m,$  $ \nu^1, \dots, \nu^{n})$ are the
coordinates of $E^{m.n}$

One can consider the functional $\Phi_A(\Omega)$ given on
(m.n)--supersurfaces by the following expression:
                   $$
                \Phi_A(\Omega)=
                   \int A
                   \left(
                  z^A(\zeta),
           {\partial z^A\over\partial  \zeta^B},\dots,
 {\partial^k z^A\over\partial\zeta^{B_1}\dots\partial\zeta^{B_k}}
                  \right)
                d^{m+n}\zeta
                                                \eqno(3.1.1)
                       $$
where the function $A$ is obeyed to the following condition
                        $$
                        A
                      \left(
                      z^A,
          {\partial z^A\over\partial {\tilde {\zeta}^B}},\cdots,
 {\partial^k z^A\over\partial\tilde{\zeta}^{B_1}\dots
  \partial  \tilde{\zeta}^{B_k}}
                   \right)=
    {\rm Ber}\left({\partial\zeta\over\partial\tilde{\zeta}}\right)
                      \cdot
                        A
                      \left(
                        z^A,
      {\partial z^A\over\partial\zeta^B},\cdots,\
{\partial^k z^A\over\partial\zeta^{B_1}\dots\partial\zeta^{B_k}}
                      \right)\,.
                                             \eqno(3.1.2)
                         $$
In the case if the condition (3.1.2) holds
 the functional (3.1.1) does
not depend on the choice of parametrization $z(\zeta)$ of the
supersurface $\Omega$.

The function $A$ obeying the condition (3.1.2) is called (m.n)
density of rank $k$.

The (m.n) density $A$ defines the functional $\Phi_A(\Omega)$ on
(m.n) surfaces obeying to additivity condition
                  $$
\Phi_A(\Omega_1+\Omega_2)=\Phi_A(\Omega_1)+\Phi_A(\Omega_2)\,.
                                         \eqno(3.1.3)
                   $$
The densities are the most general object of integration over
surfaces [28].

Let us consider in a more details the case where the rank of
density is equal one:
                             $$
                              A=
  A\left(z^A,{\partial z^A\over\partial\zeta^B}\right)\,.
                                                \eqno(3.1.4)
                             $$

 The condition (3.1.2) can be rewritten in a following way
                        $$
          A\left(z^A, K^B_{B^\prime}
       {\partial z^A\over\partial  \zeta^B}\right)=
                    {\rm Ber}\,
                         K
                       \cdot
    A\left(z^A,{\partial z^A\over\partial  \zeta^B}\right)\,.
                                                        \eqno(3.1.5)
                           $$
                        $$
                    {\rm  Ber}
                    \pmatrix
                 {A&B\cr C&D\cr}
                        =
           {{\rm Det} (A-BD^{-1}C)\over {\rm Det}\, D}
                                        \eqno(3.1.6)
                         $$
is superdeterminant of the matrix.

In the bosonic case (if there are no odd variables) it is easy to
see that the densities which are linear functions on the
${\partial z^A\over\partial\zeta^B}$ are in one--one
correspondence with differential forms: to
$k$--form $\omega=\omega_{{i_1}\dots {i_k}}
 dz^{i_1}\wedge\dots\wedge dz^{i_k}$ corresponds
density
                        $$
                 A_\omega=
   <{\partial z^{i_1}\over\partial \zeta^1},
                   \dots,
  {\partial   z^{i_k}\over\partial \zeta^k}\,,\,\omega>=
         k!\omega_{{i_1}\dots {i_k}}
         {\partial z^{i_1}\over\partial\zeta^1}\dots
      {\partial  z^{i_k}\over\partial  \zeta^k}\,,
                        $$
                        $$
         \Phi_A(\Omega^k)=\int_{\Omega^k}\omega\,.
                                             \eqno (3.1.7)
                        $$

The equation (3.1.5) holds because  Det
 (Ber $\rightarrow$ Det
in bosonic case) is polylinear antisymmetric function on tangent
vectors ${\partial x^a\over\partial\zeta^b}$.

In the case if the density $A$ corresponds to differential form
 by (3.1.7) then Stokes theorem is obeyed
                        $$
              \Phi_{A\omega}(\partial  \Omega)=
              \Phi_{A_{d\omega}}(\Omega)
                                                \eqno(3.1.8)
                        $$
One can show that {\it in bosonic case
the densities obeying to Stokes theorem
correspond to differential forms.}

What happens in supercase?

In the bosonic case differential forms are simultaneously the
linear functions on tangent vectors on which exterior
differentiation operator can be defined and on other hand
 they are integration object (3.1.5)

In the supercase the differential form $\omega$ can be defined as
the function linear on tangent vectors which is
superantisymmetric:
                        $$
        \omega(\dots, {\bf u},{\bf v},\dots)=
             -\omega(\dots, {\bf v},
            {\bf u},\dots) (-1)^{p(u)p(v)}\,.
                                                \eqno (3.1.9)
                        $$

 In supercase (3.1.9) is not in accordance with
(3.1.6)---to differential form (3.1.7) does not correspond density.

One have to construct the right generalization of differential
form (considering as integration object),
 so called psendodifferential forms as a density obeying to Stokes
theorem. It is the way which was at beginning developed in
[19,20] and was studied in general case in [2,3,62--68].

For defining pseudodifferential forms we have to check the
conditions which one have to put on the density
 (3.1.4) for having the
Stokes theorem (3.1.8) (see for details [64]).

Let two (m.n) surfaces $\Omega_0$ and $\Omega_1$ are given by
parametrization $z^A_0=z^A_0(\zeta^B)$ and
 $z^A_1=z^A_1(\zeta^B)$
correspondingly and
                $$
z^A=z^A(t,\zeta^B)\,,\,(0\leq t \leq 1):
             \quad
z(0,\zeta^B)=z_0(\zeta^B),\,
z(1,\zeta^B)=z_1(\zeta^B)
                                        \eqno(3.1.10)
                 $$

is a parametrization of $(m+1.n)$ surface ${\cal V}$
                 $$
\partial  {\cal V}=\Omega_1-\Omega_0
                                        \eqno (3.1.11)
                  $$
 (up to a boundary terms)
Then if $A$ is a density of rank 1 we have
                 $$
       \Phi_A(\partial  {\cal V})=
     \Phi_A(\Omega_1)-\Phi_A(\Omega_0)=
                  \int
                  d\zeta^{m+n}
                \int^{1}_{0}dt
                {d\over dt}
                    A
                  \left(
  z^A(t,\zeta^B),{\partial z^A(t,\zeta^B)\over\partial\zeta^B}
                  \right)=
                    $$
                    $$
                   \int
                d\zeta^{m+n}
                \int dt
               \left[
               \left(
               {dz^A\over dt}\,
      {\partial  A\over\partial  z^A}+
              {dz^A_B\over dt}\,
         {\partial  A\over\partial  z^A_B}
                 \right)
                 \right]=
                        $$
                        $$
                        \int
                     d\zeta^{m+n}
                     \int dt
                     \left[
                   {dz^A\over dt}\,
         {\partial  A\over\partial  z^A}+
                 {d\over d\zeta^B}
                     \left(
                     {d  z^A\over dt}\,
                  {\partial A\over\partial z^A_B}
                      \right)
                         -
                  {dz^A\over dt}
              {d\over\partial  \zeta^B}
         {\partial  A\over\partial  z^A_B}
                (-1)^{p(B)p(A)}
                    \right]=
                       $$
                       $$
                       \eqalign
                       {
                   \int d\zeta^{m+n} dt
                         (
                   {dz^A\over dt}
         {\partial  A\over\partial  z^A}-
                        (
                  {dz^A\over dt}
        {\partial z^{A^\prime}\over\partial\zeta^B}
      {\partial^2 A\over\partial z^{A^\prime}\partial z_B^A}
                          +&
                     {dz^A\over dt}
{\partial z^{A^\prime}_{B^\prime}\over\partial\zeta^B}
{\partial ^2 A\over\partial  z^{A^\prime}_{B^\prime} \partial  z^A_B}
                         )
                 (-1)^{p(A)p(B)}
                        )
                        \cr
                        +&
           {\rm  boundary\, terms\,.}
                         }
                                                \eqno (3.1.12)
                         $$
  (We use notation $z^A_B={\partial z^A\over\partial \zeta^B}$).

{}From (3.1.11), (3.1.12) one can see that if the last
 term in integral
(3.1.12) vanishes:
                         $$
                   {dz^A\over dt}
              {\partial^2 z^{A^\prime}
                       \over
       \partial  \zeta^B\partial  \zeta^{B^\prime}}
                 {\partial ^2 A
                        \over
     \partial  z^{A^\prime}_{B^\prime}\partial  z^A_B}
                     (-1)^{p(A)p(B)}=0\quad {\rm i.e}
                           $$
                            $$
                {\partial^2 A
                     \over
 \partial   z^{A^\prime}_{B^\prime}\partial  z^A_B}=
      -(-1)^{p(B)p(B^\prime)+(p(B)+p(B^\prime))p(A)}
               {\partial ^2 A
                      \over
 \partial   z^{A^\prime}_B\partial  z^A_{B^\prime}}
                                        \eqno (3.1.13)
                        $$
then this integral can be considered as $(m+1.n)$ density $dA$ of
rank 1. The differential is defined by the relation
                   $$
                   dA
                   \left(
          z^A,{\partial z^A\over\partial  \zeta^B},\,
              {dz^A\over dt}
                \right)=
               {d z^A\over dt}\,
        {\partial A\over\partial z^A}-
             {dz^A\over dt}\,
             {\partial z^{A^\prime}\over\partial\zeta^B}\,
 {\partial ^2 A\over\partial z^{A^\prime}\partial z^A_B}
                (-1)^{p(A)p(B)}\,.
                              \eqno(3.1.14)
                     $$
We come to correct definition (3.1.14) of the exterior differential
$d$ of the density of rank 1 in supercase if the condition (3.1.13)
holds (see for details [Vor]).(Of course in usual case from
(3.1.13)) immediately follows the statement after (3.1.8)).

The density is called pseudodifferential form if condition
(3.1.13) holds.

 If $A$ is pseudodifferential form then
$dA$ is pseudodifferential form too.
\medskip
{\bf Example 3.1.1.}

In  the superspace $E^{M.N}$
 with coordinates
 $z^A=(x^1,\dots,x^M,\theta^1,\dots,\theta^N)$
 we consider $(m.n)$ density of rank 1.
                        $$
                    A={\rm Ber}
                    \left(
            {\partial  z^A\over\partial \zeta^B}
                 L^{B^\prime}_A
                       \right)
                                        \eqno (3.1.15)
                           $$
Where $\zeta^B$ are coordinates of $E^{m.n}$,
$L_A^B $ is $(m.n)\times(M.N)$ arbitrary matrix.

$A$ is density because condition (3.1.2) is evidently satisfied.

Indeed (3.1.15) is pseudodifferential form. The condition
(3.1.13) can be checked by straightforward but long computations.
(Alternatively (3.1.13) for (3.1.15) follows from the fact that
(3.1.15) is proportional to volume form on $E^{m.n}$. The volume
form evidently obeys to (3.1.13) because the conditions
(3.1.13) are reparametrization invariant).

It is useful to consider two particular cases of (3.1.15).

a) $n=0\,\, (L^{B^\prime}_A)=0$ if $p(B^\prime)=1)$. In this case
${\rm  Ber}\rightarrow {\rm  Det}$ and to $A$ corresponds
differential form
                        $$
                     \omega_A=
            L_{A_1}^1\dots L^m_{A_m}
            dz^{A_1}\wedge\dots\wedge dz^{A_m}
                                                \eqno(3.1.16)
                         $$

b) $m=n=1$ and $L^B_A$ is such that
                $$
            A={\rm Ber}
{\partial (x^1,\theta^1)\over\partial (\xi,\eta)}=
{x_\xi^1\over\theta^1_\eta}-
  {x^1_\eta\theta^1_\xi\over(\theta^1_\eta)^2}
                                        \eqno(3.1.17)
              $$

$z^A=(x^1,\dots,x^M, \theta^1,\dots,\theta^N)$,
 $\zeta^B=(\xi,\eta)$ ($\zeta$ is even and $\eta$ is odd.)

(3.1.17) is the simplest example of non--linear
pseudodifferential form.
               \medskip
In the [2,3] Baranov and Schwarz suggested the following
construction producing the pseudodifferential form which seems
natural in spirit of ghost technique.

For $(M.N)$ dimensional superspace $E^{M.N}$ let $STE^{M.N}$ be a
superspace associated with tangent bundle $TE^{M.N}$ of the
superspace  $E^{M.N}$. (If $z^A$ are coordinates on $E^{M.N}$
then  $(z^A, z^{*A})$ are coordinates  on $STE^{M.N}$
where $z^{*A}$ transform as $dz^A$ and have reversed parity
                                $$
                        p(z^{*A})=p(z^A)+1
                                $$

  The superspace $STE^{M.N}$ have dimension $(M+N.M+N)$.

 Then arbitrary function \footnote{*} {The function W have to obey the
  conditions on infinity by even variables for (3.1.18) being correct}
 $W(z,z^*)$ on  $STE^{M.N}$  defines $(m.n)$ density of rank 1
                         $$
                        A_W
                     \left(
        z^A,{\partial  z^A\over\partial  \zeta^B}
                     \right)=
                     \int
                       W
                       \left(
                    z^A,z^{*A} =
     \nu^B {\partial z^A\over\partial \zeta^B}
                \right)
                  d\nu^{n+m}
                                                \eqno(3.1.18)
                          $$
where $\nu^B$ have reversed parity to $\zeta^B$
                        $$
              p(\nu^B)=p(\zeta^B)+1.
                                                \eqno(3.1.19)
                       $$
Indeed it is easy to check using (3.1.19) that (3.1.18) obeys to
condition (3.1.2)
                      $$
                   A_W
                   \left(
        z^A,K^B_{B^1}
        {\partial z^A\over\partial \zeta^B}
                \right)=
                  \int
                    W
                  \left(
              z^A,\nu^{B^\prime}K^B_{B^\prime}
           {\partial z^A\over\partial \zeta^B}
                \right)
                  d\nu^{n+m}=
                  $$
                  $$
                \int
                  W
                \left(
           z^A,\tilde{\nu}^B
           {\partial z^A\over\partial \zeta^B}
                  \right)
                    d\nu^{n+m}=
                   {\rm Ber}
                  (K^ B_{B^\prime})
                       \int
                        W
                     \left(
                z^A,\tilde {\nu}^B
                   {\partial z^A\over\partial \zeta^B}
                     \right)
                    d\tilde{\nu}^{n+m}
                                            \eqno(3.1.20)
                      $$
One can easy check by direct computation that the density $A_W$
in (3.1.18) obeys to condition (3.1.13).

We say that the function $W$ is BS representation of the
pseudodifferential form $A$.

One can see by straightforward calculations that to the exterior
differential $d$ of the pseudodifferential form $A$
 (See (3.1.14)) corresponds the operator
                    $$
            {\hat d}=z^{*A}
            {\partial\over\partial z^A}
                                                \eqno (3.1.23)
                     $$
in BS representation. Namely it can be checked using (3.1.14) and
(3.1.18) that
                       $$
                A_{{\hat d}W}=\pm d(A_W)
                                                \eqno (3.1.24)
                       $$
if $A_W$ is $(m.n)$ density and $A_{{\hat d}W}$ is
$(m+1.n)$ density.

{\bf Example 3.1.2}

Let us consider the function
                  $$
                  W=
          {1\over\sqrt{\pi}}
        e^{-(x^{1*})^2}\theta^{1*}
                                                \eqno (3.1.25)
                        $$
on the $STE^{M.N}$

($z^A=(x^1,\dots, x^M, \theta^1,\dots, \theta^N)$,
$z^{*A}=(x^{*1},\dots, x^{*M}, \theta^{*1},\dots, \theta^{*N})$.)

The (1.1) density corresponding to (3.1.25) by (3.1.18)
                        $$
                      A\left(
     z, {\partial z\over\partial\zeta},
     {\partial z\over\partial\theta}
            \right)=
            {1\over\sqrt{\pi}}
                \int e^
     {-(\nu x^1_\zeta+tx^1_\eta)^2}
                \left(
    \nu\theta^1_\zeta+t\theta_\eta^1
               \right)
               d\nu dt=
       {x^1_\zeta\over\theta^1_\eta}-
        {x^1_\eta\theta^1_\xi\over(\theta^1_\eta)^2}
                            \eqno (3.1.26)
                     $$
coincides with the density (3.1.17).
 To generate the density (3.1.15) from Example 3.1.1
  by the construction (3.1.18)one can consider
  instead (3.1.25) the following formal expression
                 $$
          W=\int e^{z^{*A} L^B_A c_B}dc
                              \eqno(3.1.27)
                 $$
                 where $p(c_B)=p(\zeta^B)+1$.
       (Compare with (2.3.3))

It is easy to see that formally (3.1.27) gives (3.1.15).
 But it have sense only in the case where $L^B_A=0$ if
  $p(B)=1$ .(See footnote before eq. (3.1.18)).
                  \medskip
For us it is most interesting the case where pseudodifferential
form is closed---i.e. the density obeys to condition (3.1.13) and
                       $$
                dA=0 \quad{\rm i.e.}\quad
       {\partial A\over\partial z^A}-
   {\partial z^{A\prime}\over\partial \zeta ^B}
   {\partial^2 A\over\partial z^{A\prime}\partial z^A_B}
            (-1)^{p(A)p(B)}=0
                                        \eqno(3.1.28a)
                     $$
or   in BS representation
                $$
              \hat{d}W_A=
     z^{*A}{\partial W\over\partial z^A}=0
                                        \eqno (3.1.28b)
                  $$
In other words condition of closeness means that Euler--Lagrange
equations of the functional (3.1.1) are trivial [64].

It is these densities which arise when we reduce the partition
function integral (2.2.2) to the integral over the surfaces in the
space of field configurations defined by gauge conditions. The
gauge independence of this integral means that the corresponding
density is closed. But in field theory this surface is defined
not by parametrization but by equations ("gauge conditions")
  We need to consider corresponding integration
   objects.
               $$ $$
\centerline{3.2 Dual densities and closed pseudointegral forms}
                    \bigskip

The surfaces in the superspace can be defined not by
parametrization, but by dual construction---by equations.

Let $\Omega^ {(m.n)}$ be a $(m.n)$ dimensional supersurface in the
superspace $E^{M.N.}$ defined by equations
                        $$
                     F^a (z^A) =0
                                                \eqno (3.2.1)
                        $$
Where $F^a=(f^1,\dots,f^{M-m},\varphi^1,\dots,\varphi^ {N-n})$
 are coordinates of the superspace $E^{M-m.N-n}$ ($f$ are even,
 $\varphi $  are odd).

Let
                                $$
                dv=\rho (z)dz^1\dots dz^n
                                                \eqno(3.2.2)
                               $$
      be a volume form on $E^{M.N}$
Then (3.1.1) can be replaced by the functional:
                          $$
                \Phi_{\tilde A}(\Omega)=
                     \int
                   \tilde{A}
                   \left(
                   z^A,
 {\partial F^a(z)\over\partial z^A},\dots,
  {\partial ^k F(z)\over\partial z^{A_1}\dots\partial z^{A_k}}
               \right)
                \prod_a
               \delta
               (F^a)dv.
                             \eqno(3.2.3)
                   $$
   where $\tilde{A}$ is obeyed to the condition
                     $$
           \tilde{A}
                \left(
                  z^A,
           {\partial F^{\prime^c}(z)\over\partial z^A},\dots,
{\partial ^k F^{\prime c}(z)\over\partial z^{A_1}\dots\partial z^{A_k}}
                \right)=
        {\rm Ber}\left( \eta ^c_d\right)
              \tilde{A}
               \left
               ( z^A,
      {\partial F^c(z)\over\partial z^A},\dots,
{\partial ^k F^c(z)\over\partial z^{A_1}\dots\partial z^{A_k}}\right)
                                                \eqno(3.2.4)
                        $$
 ($F^{\prime c}(z)=\eta ^c_d F^d(z)$ determine the same
surface $\Omega ^{m.n}$).

$\tilde{A}$ is called $(m.n)$ $D$--density
 (dual density) of rank $k$ [28]. It is easy to
see that in the same way like for usual densities, if conditions
(3.2.4) hold then (3.2.3) does not depend on the choice of the
functions $\left\{F^a(z)\right\}$ defining the surface $\Omega$
by the equation (3.2.1).

$D$--density $\tilde{A}$ corresponds to density $A$ if for
arbitrary surface $\Omega $
                   $$
     \Phi_{\tilde{A}}(\Omega)=\Phi_A(\Omega )
                                                \eqno(3.2.5)
                   $$

More precisely if the surface $\Omega ^{m.n}$ is given by
equation
                      $$
         F_0^a(z^A)=y^a-r^a(s^B)
                        $$
and its parametrization by
                         $$
    z_0^A(\zeta):\quad  y^a(\zeta^B)=r^a(\zeta^B),\,
              s^B(\zeta)=\zeta^B
                  \qquad (z^A=(y^a,s^B))
                                        \eqno(3.2.6)
                              $$
               then the $\tilde A$ corresponds to $A$ if
                 $$
            {\tilde A}
            \left(
  z^A,{\partial F_0^a\over\partial z^A},\dots,
  {\partial^k F_0^a\over\partial z^{A_1}\dots\partial z^{A_k}}
               \right)=
                     A
              \left(
   z^A_0(\zeta),
   {\partial z_0^A\over\partial\zeta ^B},\dots,
 {\partial ^k z_0^A\over\partial z^{B_1}\dots\partial \zeta ^{B_k}}
             \right)
                                 \eqno(3.2.7)
                       $$
 in the case $\rho=1$ in the (3.2.2).

In the next section we consider the examples of $D$--densities
arising in odd symplectic geometry.

If the density $A$ corresponds to differential form
$ w_{i_1\dots i_k}dx^{i_1}\wedge\dots \wedge dx^{i_k}$
 in the space $E^n$ (in bosonic case) then it is easy to see that dual density
 $\tilde{A}$ corresponds to integral form
                       $$
               W^{{i_1}\dots i_{n-k}}
               {\partial \over\partial x^{i_1}}
                \wedge\dots\wedge
      {\partial \over\partial x^{i_{n-k}}}
                           $$
     such that
$W^{i_1\dots i_{n-k}}={1 \over\rho }
 \epsilon^{i_1\dots i_{n-k}
  j_1\dots j_k}\omega_{j_1\dots j_k}$ and
                $$
            \tilde{A}=
            W^{i_1\dots i_{n-k}}
     {\partial f^1\over\partial x^{i_1}},\dots,
     {\partial f^{n-k}\over\partial x^{i_{n-k}}}
                                        \eqno(3.2.8)
                $$
where $\rho$ is given by volume form (3.2.2).

To construct $D$--densities which are dual to
pseudodifferential forms (so called pseudointegral forms) one
have to check the eq. (3.1.8) on the language of $D$
--densities.

We represent only the final results of these calculations:

Let $A(z^A,{\partial F^a\over\partial z^A} )$
be a $D$--density of rank 1.
                        $$
                        A
                      \left(
                  z^A,K^a_b
      {\partial F^b\over\partial z^A}
                   \right)=
                    {\rm Ber} K\cdot
                        A
                  \left(
                   z^A,
          {\partial F^a\over\partial z^A}
                  \right)
                                                \eqno(3.2.9)
                        $$
This density is closed if
                        $$
{\partial^2 A\over\partial F^a_A \partial  F^b_B}
            =-(-1)^
            {p(A)p(B)+(p(A)+p(B))p(a)}
 {\partial^2 A\over\partial F^a_B \partial  F^b_A}
                                              \eqno(3.2.10)
                          $$
(Compare with (3.1.13)) and
                         $$
                      (-1)^{p(a)p(A)}
                      {1\over\rho}
                      {\partial\over\partial z^A}
                         \left(
                           \rho
               {\partial A\over\partial F^a_A}
                     \right)=
                        0.
                                           \eqno(3.2.11)
                        $$
One can come to (3.2.10), (3.2.11) considering the variation of functional
(3.2.3) under the infinitesimal variation of surface $\Omega $
(Compare with (3.1.12)).

Analogously to (3.1.18) one can develop Baranov--Schwarz
 procedure for pseudointegral forms [40].

For $(M.N)$ dimensional superspace $E^{M.N}$ let $ST^*E^{M.N}$ be
a superspace associated with cotangent bundle $T^*E^{M.N}$ of the
superspace $E^{M.N}$. $(z^A,z^*_A)$ are the coordinates of
$T^*E^{M.N}$, $z^*_A$ transform as
${\partial\over\partial z^A}$ and
have reversed parity
                $$
        p(z^*_A)=p(z^A)+1.
                $$
Then the arbitrary function $W(z,z^*)$ on $T^*E^{M.N}$ (see the
footnote before (3.1.18)) defines $D$--density of rank 1:
                  $$
                A \left(
                z^A,
          {\partial F^a\over\partial z^A}
              \right)=
                \int W
                \left(
          z^A,z^*_A=\nu_a{\partial F^a\over\partial z^A}
                 \right)
                   \prod_a d\nu_a,
                                        \eqno(3.2.12)
                    $$
                    $$
             p(\nu_a)=p(F^a)+1 .
                    $$

(Compare with (3.1.18)).

$A$ indeed is density (The condition (3.2.9) is evidently
satisfied for (3.2.12) as in (3.1.20)).
The conditions (3.2.10) can checked by
direct computation.

Comparing (3.2.11) with (3.2.12) one can see that the density
(3.2.12) is closed if
                        $$
           {1 \over\rho }
           {\partial \rho \over\partial z^A}
           {\partial W\over\partial z_A^*}+
           {\partial ^2 W\over\partial z^A\partial z_A^*}=0.
                                \eqno(3.2.13)
                        $$
In the 4--th Section we give the interpretation of (3.2.13) in
the terms of odd symplectic geometry.---Indeed this formula is
strictly connected with BV master---equation.

Now let us consider example which we will use later:

{\bf Example 3.2.1}

Let $\{ R^A_a\}$ be a set of vector fields on
$E^{M.N}$. One can consider $D$--density---
pseu\-do\-in\-teg\-ral form:
             $$
        \tilde{A}
        \left(
        {\partial F^a\over\partial z^A}
          \right)=
 {\rm Ber}\left({\partial F^a\over\partial z^A}R^A_b\right)\,.
                                        \eqno (3.2.14)
              $$
(Compare with (3.1.15)).

It is the density which arise in (2.3.2).

One have to note that
$(m.N)$  $D$--density (of maximal odd dimension)
are polynomial by ${\partial F^a\over\partial z^A}$.
It is just Bernstein---Leites integral forms [19,20] (see also [28]).

The $D$--density (3.2.14) formally generates by the
function
                $$
                W
                \left(
                z^A,z_*^A
                \right)=
                \int e^
          {c^a R^A_a z_A^*}dc
                             \eqno(3.2.15)
                  $$
                  $$
         (p(c^a)=p(a)+1)
                  $$
The equation (3.2.15) is correct in the case if all vector fields
$R^A_a {\partial \over\partial z^A}$ are even.

 $c^a$ in (3.2.15) are nothing that usual ghosts in
 Faddeev---Popov trick (Compare with (2.3.3))

\magnification=1200
\def\p{\partial}
\def\al{\alpha}
\def\ep{{\cal E}}
\def\ph{\varphi}

        \centerline  {\bf 4  Odd Symplectic Geometry }
            \bigskip
           \centerline  { 4.1  Basic Definitions }
                \medskip
    Let $M^{2n}$ be an $2n$--dimensional manifold provided
    with closed non--degenerated two form $w$:
                      $$
                 dw=0\,,\quad {\rm Det}w_{ij}\neq 0\,,
                                                     \eqno(4.1.1)
                      $$
 where $w=w_{ij}dx^i\wedge dx^j$ in the local coordinates
  $(x^1,\dots,x^{2n})$.

  The pair $(M^{2n},\, w)$ is called symplectic manifold.

   The non--degenerated two--form (4.1.1) establishes
   one--one correspondence between $TM$ and $T^*M$:
                       $$
   \forall {\bf u}\in T_m M\, \,(m\in M)
                   \rightarrow\,\,{\rm one}\,{\rm form}\,
                   w_m\in T^*_m M:\,
           \forall {\bf \xi}\in T_m M\,
           w_m({\bf \xi})=w({\bf \xi},{\bf u})\,.
                                                          \eqno (4.1.2)
                        $$
  According to (4.1.2) to every function $f$ on $M$ corresponds
  a vector field ${\bf D}_f$ which in coordinates is
                        $$
      D^i_f\partial_i=(w^{-1})^{ij}\partial_j f \partial_i\,.
                                                           \eqno (4.1.3)
                       $$
       (To vector field ${\bf D}_f$ corresponds one form $df$ by
       (4.1.2)).

 The Poisson bracket of two functions $f$ and $g$ is equal
                       $$
        \{f,g\}=w({\bf D}_f,{\bf D}_g)=
        {\partial f\over \partial x^i}
               (w^{-1})^{ij}
          {\partial g\over \partial x^j}.
                                                 \eqno (4.1.4)
                        $$
  It obeys to Jacoby identity
                        $$
  \{\{f,g\},h\}+{\rm cyclic\,permutation}=0
                                               \eqno (4.1.5)
                         $$
     which follows from (4.1.1).

 The group $G^{can}$ of the symplectomorphisms (canonical
 transformations) of
 $(M^{2n}\,,w)$ i.e. the diffeomorphisms preserving the two form $w$
  is infinite--dimensional.---
 To every function (Hamiltonian) $H$ corresponds
 infinitesimal transformation ${\bf D}_H$
                         $$
  {dx^i\over dt}=D_H^i=\{H,x^i\}\,,\quad
       {\cal L}_{{\bf D}_H}w=0\,.
                                           \eqno (4.1.6)
                          $$
   There exists unique (up to multiplication on constant)
   $2k$ density which is $G^{can}$--invariant.
    (We say that a density $A$ on $E$ is invariant under the action of
    a group $G$ of transformations of $E$ if in (3.1.1)
                         $$
         \forall g\in G\,,\quad
                 \Phi_A(\Omega^g)=\Phi_A(\Omega)\,.)
                                          \eqno (4.1.7)
                           $$

   It is closed density which corresponds to
   $k$--times wedged product of the form $w$ :
                            $$
           \Phi_A(\Omega) =
           \int_{\Omega}w\wedge\dots\wedge w
                                              \eqno (4.1.8)
                           $$
  It is  a well--known Poincare--Cartan integral invariant
of canonical transformations [1].

 The integrand in (4.1.8) is $G^{can}$ invariant closed
  $2k$--density
  of rank 1
                           $$
      A(x^i,{\partial x^i\over\partial\xi^\alpha}) =
                        \sqrt
                          {
                       {\rm Det}
                        \left(
            {\partial x^i\over\partial\xi^\alpha}
                         w_{ij}
            {\partial x^i\over\partial\xi^\beta}
                         \right)
                            }
                                               \eqno (4.1.9)
                             $$
   where $x^i(\xi^\alpha)$ is the parametrization of surface $\Omega$.

The dual D--density ${\tilde A}$ corresponding to $A$ is
                           $$
                      {\tilde A}
          (x^i,{\partial f^a\over\partial x^i}) =
                        \sqrt
                          {
                          Det
                        \left(
                        \{f^a,f^b\}
                         \right)
                            }
                                               \eqno (4.1.1.9a)
                             $$
  where the equations $f^a=0$  define the surface $\Omega$ .

     In the case $k=n$ the density (4.1.9) is $G^{can}$--invariant
     volume form corresponding to the symplectic structure.

 Locally there exist coordinates in which the form $w$
 (4.1.1) defining symplectic structure have canonical form
 (Darboux Theorem):
                    $$
                w=\sum_{i=1}^n
                    dx^i
                   \wedge
                    dx^{i+n}\,.
                                          \eqno(4.1.10)
                     $$

    {\bf Remark}.
   Indeed one can prove more :  By suitable canonical
  transformation one can make "flat" a surface in a vicinity
  of arbitrary point (for any $\xi_0$ the derivatives
  ${\partial^k x^i\over \partial\xi^{[k]}}\vert_{\xi_0}$
   for $k\geq 2$ can be cancelled by suitable
   canonical transformation). From this fact in
     particularly follows that
       the density (4.1.9) is a unique
  $G^{can}$--invariant density in the class of densities
  of arbitrary rank $k$ [41].

  The symplectic geometry in pure bosonic case in contrary
  to Riemannian one is "poor"
  because the group of transformations is "rich" .
  In Riemannian geometry there are the invariant
  densities of higher degrees constructed  via
  the curvature and its covariant derivatives.---
  The analogue of (4.1.9) which is a volume form
   of the surface
  ($w_{ij}\rightarrow g_{ij}$ in (4.1.9)) is not
  unique invariant density.
    \medskip

   Now we represent the superizations of the constructions above.

     Let
                          $$
                      w=w_{AB}dz^A dz^B
                                            \eqno (4.1.11)
                           $$
 be closed non--degenerated two form on the superspace
    $E^{M.N}$ with coordinates

    $z^A=(x^1,\dots,x^M,\theta^1,\cdots,\theta^N)$.

 $w$ is a linear superantisymmetric function on tangent vectors:
                           $$
                           \eqalign
                           {
     w({\bf u},{\bf v}\lambda)&= w({\bf u},{\bf v})\lambda\,,
                       {\rm if}\,\lambda\in {\bf R}\cr
                     w({\bf u},{\bf v})& =
                   - w({\bf v},{\bf u})
                        (-1)^
                  {p({\bf u})p({\bf v})} .
                            }
                                                   \eqno(4.1.12)
                            $$
  In coordinates
                            $$
                            \eqalign
                            {
                   w_{AB}&=(-1)^{1+p(A)p(B)}w_{BA}\,,\cr
                   w(u^A\partial_A,v^B\partial_B)&=
                       u^A w_{AB}v^B
                            (-1)^
               {(p(u)+p(A))p(w)+(p(v)+p(B))p(B)}.
                            }
                                                       \eqno(4.1.13)
                              $$
    The closeness condition $dw=0$ is
                               $$
                          (-1)^{p(A)p(C)}
                          \partial_A w_{BC}
                                 +
                      {\rm cyclic \, permutation}
                                  \,
                                  =
                                   0\,.
                                                      \eqno(4.1.14)
                                $$

     The non--degeneracy of $w$
     (i.e. the matrix $w_{AB}$ is invertible)
     on $E^{M.N}$ means that
     $M$ is even if $w$ is even and
     $M=N$ if $w$ is odd.

  The analogue of Darboux Theorem [57] states that
  there exist coordinates (Darboux coordinates) in which
  the two form $w_0$ defining an even symplectic
  structure on $E^{2M.N}$ have the following
  canonical form:
                     $$
                    w_0=
                   \sum_{i=1}^M
                dx^i\wedge dx^{i+M}+
                 \sum_{\alpha=1}^N
                  \epsilon_\alpha
                  (d\theta^\alpha)^2\,,
                   \quad    (\epsilon_\alpha =\pm 1)
                                             \eqno(4.1.15)
                      $$
 and the two form $w_1$ defining an odd symplectic
  structure on $E^{M.M}$ have the following
  canonical form:
                       $$
                    w_1=
                   \sum_{i=1}^M
                dx^i\wedge d\theta^i\,.
                                             \eqno(4.1.16)
                      $$
 (On $E^{2M.2M}$ one can consider two simplectic structures
 of the different grading simultaneously ( see [34,37,38,47])).

  Using  (4.1.13) and (4.1.14) one can establish a
  superversion of the equations (4.1.3) and (4.1.4):
                        $$
                     D_f^A\partial_A=
                     (w^{-1})^{AB}
                      \partial_B f
                         (-1)^
             {(p(f)+p(w)+p(B))p(B)}\partial_A
                                            \eqno (4.1.17)
                         $$
    and formulae for Poisson bracket:
                         $$
                       \{f,g\}=
                {\partial f\over\partial z_A}
                    (w^{-1})^{AB}
                {\partial g\over\partial z_B}
                         (-1)^
                     {(p(f)+p(w))p(A)}\,.
                                                \eqno(4.1.17a)
                          $$
    For computing (4.1.17a) we have to note that the
    inverse matrix in the superspace has the inverse parity:
                        $$
    (w^{-1})^{AB}=(-1)^{(p(A)+1)(p(B)+1)}(w^{-1})^{BA}.
                        $$
     (Compare with (4.1.13)).
   In Darboux coordinates on $E^{2M.N}$ the even
   Poisson bracket corresponding to (4.1.15) have the form:
                           $$
                         \{f,g\}=
                         \sum_{i=1}^M
                         \left(
                   {\partial f\over \partial x^i}
                 {\partial g\over \partial x^{i+M}}
                              -
                 {\partial f\over \partial x^{i+M}}
                   {\partial g\over \partial x^i}
                           \right)
                             +
                       \sum_{\alpha=1}^N
                         \epsilon_\alpha
                           (-1)^{p(f)}
                  {\partial f\over \partial \theta^\alpha}
                   {\partial g\over \partial \theta^\alpha}
                                                          \eqno (4.1.18)
                             $$

 and  in Darboux coordinates on $E^{M.M}$ the odd
   Poisson bracket corresponding to (4.1.16) have the form:
                           $$
                         \{f,g\}_1=
                         \sum_{i=1}^M
                         \left(
                   {\partial f\over \partial x^i}
                 {\partial g\over \partial \theta^i}
                               +
                              (-1)^
                              {p(f)}
                {\partial f\over \partial \theta^i}
                   {\partial g\over \partial x^i}
                              \right)\,.
                                                           \eqno(4.1.19)
                             $$
     The Hamiltonian mechanics can be formulated
      in the terms of even as well as odd symplectic
  structures [43,60,36].
      \medskip
       The formulae above are similar for even and odd structures.
        But there is essential difference between these
        structures.

  An even symplectic structure is nothing but natural lifting
  on $E^{2M.N}$ of the symplectic structure of the underlying
  space $E^{2M}$. And it is natural that it is very similar
  to symplectic structure in pure bosonic case.

 For example the changing Det$\rightarrow Ber$ in
 (4.1.9,4.1.9a) leads to straightfor\-ward ge\-ne
 ra\-li\-za\-tion of
 Poincare--Cartan invariant on the
  supercase (if the structure is even) [41]:
                       $$
      A(z^A,{\partial z^A\over\partial\zeta^\alpha}) =
                        \sqrt
                          {
                       {\rm Ber}
                        \left(
            {\partial z^A\over\partial\zeta^\alpha}
                         w_{AB}
            {\partial z^A\over\partial\zeta^\beta}
                         \right)
                            }
                                               \eqno (4.1.20)
                             $$
   and corresponding D--density:
                            $$
                      {\tilde A}
          (z^A,{\partial F^a\over\partial z^A}) =
                        \sqrt
                          {
                        {\rm Ber}
                        \left(
                        \{F^a,F^b\}
                         \right)
                            }   .
                                                  \eqno (4.1.21)
                             $$

  Of course in the supercase the invariant density
  cannot be anymore represented as integrand in (4.1.8) because
  form is not anymore integration object.(See section 3.1). But
  one can show that as well as in bosonic case (4.1.8)
  the density (4.1.20) is closed and there is no invariants
  in higher derivatives [41,2,3].
  (The Remark above is valid in this case too.)

 It is not the case for odd symplectic geometry. At first
 its ancestor in pure bosonic case is Lie derivative construction,
 not the symplectic geometry:

  {\bf Example 4.1.1}

  Let $E^{M.M}=ST^*E^M$ be a superspace associated with
  cotangent bundle of the space $E^M$. (See subsection 3.2).

 $ST^*E^M$ is naturally provided with odd
 symplectic structure
                  $$
        w_1 =dx^i d\theta_i,\quad (\theta_i=x^*_i).
                                                   \eqno(4.1.22)
                  $$
 Then to vector field ${\bf R}=R^i(x)\partial_i$ on $E^M$
 corresponds the function $R=R^i(x)\theta_i$ and
                  $$
          {\cal L}_{{\bf R}}f=
     R^i(x){\partial f\over \partial x^i}=
                \{f,R\}.
                                                      \eqno(4.1.23)
                   $$
   More generally to every polyvectorial field
   ${\bf T}=T^{A_1\dots A_n}
   \partial_{A_1}\wedge\dots\wedge\partial_{A_n}$
   corresponds the function
                    $$
                    \sigma({\bf T}) =
              W_{\bf T}(x,\theta)=
                        =
         T^{A_1\dots A_n}\theta_1\dots\theta_n
                                                  \eqno (4.1.24)
                     $$
                and
                     $$
         \sigma([{\bf T}_1,{\bf T}_2])=
         \{\sigma({\bf T}_1),\sigma({\bf T}_2)\}
                                                   \eqno (4.1.25)
                      $$
         where $[\quad,\quad]$ is Schouten bracket.
         \medskip

  We see from this example that odd symplectic
  geometry is strictly connected with classical geometrical
  objects. And it is not surprising that in
  the terms of odd bracket some classical geometrical
  constructions can be formulated in a elegant way
   ([38,47--49]).

 In the next subsections we  will consider the geometrical
 constructions in odd symplectic geometry
 which have no analogues for even one and
 which play a crucial role in the formulating
 BV formalism.
    The essential difference of odd symplectic geometry
         from even one is that the transformations preserving
         odd bracket do not preserve any volume form.
         (The formulae (4.1.20,4.1.21) have not sense in the
         case if $w$ is odd.) To consider the integration theory
         we provide an odd symplectic space with
         additional structure---volume form.

     \medskip
  \centerline {4.2  $\Delta$--operator in odd symplectic geometry.}
      \medskip

   Let $E^{M.M}$ be provided with odd symplectic structure
   and with volume form $dv$
                      $$
                dv=\rho(x,\theta)
                dx^1\dots dx^M
                d\theta_1\dots \theta_M
                                             \eqno(4.2.1)
                      $$

   We suppose $z^A= (x^i,\theta_j)$ be Darboux coordinates (4.1.16)
   on $E^{M.M}$.
   We consider $E^{M.M}$ provided with a
   structure defined by a pair
   $(dv,\{\quad,\quad\})$ where $\{\quad,\quad\}$  is the
   odd Poisson bracket (4.1.19).
    $G^{can}_{dv}\leq G^{can}$ is the group of the
    transformations preserving the pair
    $(dv,\{\quad,\quad\})$. From here and later we call
    the structure defined by the pair  $(dv,\{\quad,\quad\})$
    the odd symplectic structure.

  We define the $G^{can}_{dv}$--invariant second order
  differential operator
  $\Delta$--operator [34] corresponding to the structure
  $(dv,\{\quad,\quad\})$ in the following way
                       $$
                    \Delta f=
                   {1\over 2}
           {{\cal L}_{{\bf D}_f}dv\over dv}=
                      div_{dv}{\bf D}_f.
                                      \eqno(4.2.2)
                        $$
     One can see by direct computation that
                        $$
                    \Delta f=
                    {1\over 2\rho}
             {\partial\over \partial z^A}
                    \left(
                      (-1)^{p(A)}
                         \rho
                       \{z^A,f\}
                       \right)
                                                     \eqno(4.2.3)
                         $$
    where $\rho$ defines volume form $dv$ by (4.2.1).
                           $$
                         \Delta f=
                       {1\over 2}
                       \{log\,\rho,f\}+
         {\partial^2 f\over\partial x^i\partial\theta_i}
                                                     \eqno(4.2.3a)
                          $$
       in Darboux coordinates.

    (In the case where $\{\,,\,\}$ is even Poisson bracket
    it is easy to see that the corresponding operator (4.2.2)
    is trivial: it is a first order differential
    operator which vanishes if a volume form corresponds
    to even symplectic structure by (4.1.20)).

   {\bf Example 4.2.1}
      $\rho=1$ in (4.2.1) then
                           $$
                           \Delta=
                           \Delta_0=
          {\partial^2 \over\partial x^i\partial\theta_i}
                                                      \eqno(4.2.4)
                            $$
  In this form this operator  was
  introduced by Batalin and Vilkovisky for
  formulating  master--equation [11,12] (see (2.2.6)).
  \medskip
  {\bf Example 4.2.2}
   Let $E^{M.M}=ST^*E^M$ be provided with natural symplectic
   structure (See Example 4.1.1). Let
                         $$
           dv=\rho(x^1,\dots,x^M)dx^1\dots dx^M
                                                      \eqno (4.2.5)
                       $$
 be volume form  on $E^M$.
 We consider the pair $(d{\hat v},\{\,,\,\})$ on
 $ST^*E^M$  where
                       $$
           d{\hat v}=\rho^2(x^1,\dots,x^M)
              dx^1\dots dx^M d\theta_1\dots d\theta_M
                                                      \eqno (4.2.6)
                       $$
  is the volume form on  $ST^*E^M$  and $\{\,,\,\}$
  is the Poisson bracket (4.1.19) which corresponds to natural
  symplectic structure (4.1.22.). Then using
  (4.2.3a), (4.2.6) and (4.1.24)  we see that $\Delta$--operator
  on $ST^*E^M$ corresponds to divergence on $E^M$:
                        $$
                 \Delta_{d{\hat v}}
                      \sigma({\bf T})=
                    {1\over \rho}
                    {\partial\rho\over\partial x^i}
            {\partial W_{\bf T}\over\partial\theta_i}
                            +
            {\partial^2 W_{\bf T}
            \over\partial x^i\partial\theta_i}=
                         \sigma
                     (div_{dv}{\bf T}).
                                           \eqno(4.2.7)
                          $$
       where $\sigma({\bf T})=W_{\bf T}$ is given by (4.1.24).

 Moreover comparing (4.1.24) (3.2.8) and (3.2.12) one can see
 that  $\sigma({\bf T})=W_{\bf T}$  is BS representation (3.2.12)
 of the D--density ${\tilde A}_{\bf T}$ corresponded to
 polyvectorial field ${\bf T}$ by (3.2.8). Then comparing
 (3.2.13) and (4.2.7) we see that closeness condition
 can be expressed in the terms of corresponding
 $\Delta$--operator [56], [40].
                          $$
        {\bf T}\,\, {\rm is}\,{\rm closed}\,
             \Leftrightarrow
                       \Delta \sigma({\bf T})=0.
                                           \eqno(4.2.7a)
                          $$
 This example where $\Delta$--operator corresponds
 to divergence describes
  an important but special case of the
  $\Delta$--operator (4.2.2).(See Theorem below).

 (In the examples 4.2.2 as well as in the examples
4.1.1 and 4.2.1  it was considered a case where
$E^{M.M}=ST^*E^M$ ($z^A\to x^i$ and $z^*_A\to \theta_i$).
By the slight modification of the considerations above
one can consider in these examples $E^{M.M}=ST^*E^{M-k.k}$
where $k\neq 0$.)
                          \medskip

  Using (4.2.3) one can see that $\Delta$--operator
  in general case obeys to conditions
                     $$
                  \Delta_{dv^\prime}f=
                  \Delta_{dv}f+
                    {1\over 2}
                    \{log\lambda,f\}
                                               \eqno (4.2.8)
                       $$
    and
                       $$
                  \Delta^2_{dv^\prime}f=
                  \Delta^2_{dv}f+
  \{\lambda^{-{1\over 2}}\Delta_{dv}\lambda^{{1\over 2}},f\}
                                                  \eqno(4.2.9)
                         $$
       where $dv^\prime=\lambda dv$.

  Following to A.S.Schwarz [56] we call the structure
 $(dv,\{\quad,\quad\})$ $SP$ structure if
 there exist Darboux coordinates in which
                     $$
                \Delta=\Delta_0
                                            \eqno(4.2.10)
                     $$
  i.e. $\rho=1$ in (4.2.1) (see eq.(4.2.4) in the Example (4.2.1)).
    \medskip
    {\bf Theorem} The following statements are equivalent:

 i) $(dv,\{\quad,\quad\})$  structure is $SP$ structure

 ii) The $\Delta$--operator corresponding to
  $(dv,\{\quad,\quad\})$  structure is nilpotent:
                   $$
               \Delta^2_{dv}=0
                                                 \eqno (4.2.11)
                   $$
 iii) the function $\rho$ corresponding to volume form
 $dv$ by (4.2.1) obeys to equation:
                    $$
               \Delta_0 \sqrt {\rho}=0.
                                                  \eqno (4.2.11a)
                    $$
  (This Theorem is stated in [56], [39], [30].
  The complete proof belongs to A.S.Schwarz [56])

    For example for the structure $(d{\hat v},\{\,,\,\})$
    from the Example (4.2.2) we come to i) if we choose
    coordinates on $E^M$ in which volume form (4.2.5)
    is trivial on $E^M$
    ($dv=dx^1\dots dx^M$). (The corresponding transformation of
    $\theta_i=x^*_i$ preserves symplectic structure.)
    The nilpotency condition ii) follows from the fact that
    $\Delta_{d{\hat v}}$ corresponds to divergence (4.2.7),
    The equation iii) is evidently obeyed.

 In general case ii)$\Leftrightarrow$iii) follows from
  (4.2.8, 4.2.9) and i)$\Rightarrow$ii) is evident from
  invariant definition (4.2.2). The ii)$\Rightarrow$i)
  needs a more detailed analysis.

  {\bf Remark} In the paper [Kh] where was first introduced the
  structure $(dv,\{\quad,\quad\})$ for arbitrary volume form
  $dv$, was done the false statement that every
  $(dv,\{\quad,\quad\})$  structure is $SP$ structure.

        \bigskip
     \centerline {4.3 Invariant densities in odd symplectic geometry.}
     \medskip
In contrary to even symplectic geometry where the invariant
densities are exhausted by the density (4.1.20) depended on first
derivatives, in odd symplectic geometry the situation is more
non--trivial.

On one  hand as it was mentioned above there are no $G^{can}$
--invariant densities, because the group of transformations
preserving odd symplectic structure does not
 preserve any volume form.
In the class of densities which are invariant under canonical
transformations preserving a fixed volume form $dv$ the first
non--trivial density (except the volume form
 itself) appears in a second derivatives [35].

In spite of this fact one can consider the density of rank 1 which
is naturally defined on Langrangian surfaces and does not change
 under infinitesimal transformations in the class of
  Lagrangian surfaces
  in the case if $(dv,\{\,,\,\})$--structure
is $SP$ structure [55,56,40].
 We consider now this density.

 Let a superspace $E^{N.N}$ be provided with
  a structure $(dv,\{\,,\,\})$ defined in previous subsection.

Let $\Lambda$--be Lagrangian surface in it (i.e the form $w$
defining simplectic structure  vanishes on it )
                 $$
             w\vert_\Lambda=0
                          \eqno(4.3.1)
                  $$
and $\Lambda$ is $(N-k,k)$--dimensional.

For example if $E^{N.N}=ST^*E^N$ then to every odd function
$\Psi(x)$ on $E^N$ corresponds $(N.0)$--Lagrangian
surface in $ST^*E^N$ defined by the equation
                 $$
\theta_i={\partial \Psi (x)\over\partial x^i}
                           \eqno(4.3.2)
                 $$
We consider later only the case $k=0$. (The case $0<k\leq n$ can
be received by slight modifications of corresponding formulae.
For example in (4.3.2) we come to $(N-k. k)$ dimensional
Lagrangian surface if we consider instead $E^{N.N}$
 $ST^*E^{N-k. k}$,
($x^i\rightarrow z^i$,$\theta_i\rightarrow z^*_i$).

If $\{{\bf t}_1,\dots,{\bf t}_n\}$ are the vectors tangent to
Lagrangian manifold $\Lambda$ in the point
$\lambda _0 \in\Lambda $ then we consider arbitrary
vectors $\{{\bf u}_1,\dots,{\bf u}_n\}$ such that
                   $$
     w ({\bf t}_i,{\bf u}_k)=\delta _{ik}
                                 \eqno(4.3.3)
                    $$
and define a density $A$ by equation [56]:
                     $$
    A\left(\lambda_0,{\bf t}_1,\dots {\bf t}_n\right)=
               \sqrt{dv (\lambda_0,{\bf t}_1,\dots,{\bf t}_n)}
                                  \eqno(4.3.4)
                     $$
(The volume form $dv$ is $(N.N)$ density of rank 1 on $E^{N.N}.$
(see Section 3.)

It can be proved that (4.3.4) is a density which
 does not depend on the choice of
the vectors $\{{\bf u}_i \}$ obeying to (4.3.3)
and this density
(more exactly the functional (3.1.1))
 is invariant under infinitesimal variation of the
Lagrangian surface $\Lambda$
  if the
 $(dv,\{\,,\,\})$--structure is $SP$ structure [56].
  We prove it later.

Instead (4.3.4) we consider $D$--density which is defined on all
$(N.0)$--dimensional surfaces and corresponds
 to the density (4.3.4) in the case if
 the surface is Lagrangian [40].
   Let a $(N.0)$-dimensional surface $\Omega$
   be defined in $E^{N.N}$ by  equations
                  $$
                  F^a=0,\quad
                  (a=1,\dots,N),\qquad
             (F^a\,{\rm are}\,{\rm odd}).
                                        \eqno(4.3.5)
                      $$
    One can consider [40]
                     $$
 {\tilde A}\left(z^A,{\partial F^a\over\partial z^A}\right)=
{1\over\sqrt{\rho }}
              {\sqrt{{\rm Ber}
 {\partial (G,F)\over\partial (x,\theta )}}
 \over\sqrt{{\rm Det}\{G^a, F^b\}}}
                                           \eqno(4.3.6)
                      $$

where $z^A=(x^1,\dots x^N, \theta_1,\dots \theta_N )$,
  are the coordinates in
   $E^{N.N}=ST^*E^N$, $\rho$ defines the volume form
                             $$
                            dv=
            \rho(x^1,\dots x^N,\theta^1,\dots,\theta^N)
            dx^1\dots dx^N d\theta_1\dots d\theta_N
                             $$
 and $\{ G^{a}\}
(a=1,\dots,N )$ are arbitrary even functions.

One can see  that
 (4.3.6) is
indeed $(N.0)$ D--density.
($F^a$ are odd so ${\rm Det}^{-1}\sim
{\rm Ber}$).

 Moreover the D--density (4.3.6) on the surface
  (4.3.5) and corresponding to it functional
  (3.2.3) $\Phi_A(\Omega)$ does not depend on the choice of
 the functions $\{ G^a\}$ if $\Omega$ is Langrangian
surface.
Indeed in this case the functions $F^a$ defining
$\Omega$ by (4.3.5) obey
to equation
                         $$
                  \{F^a,F^b \}\vert_{F^a=0}=0
                                       \eqno(4.3.7)
                         $$
    (Compare with (4.3.1)).

Let
                   $$
        \tilde{G}^a=\tilde{G}^a
        (G^1,\dots,G^N,F^1,\dots F^N)
                                          \eqno(4.3.8)
                    $$
  be another set of even functions $\{{\tilde G}^a\}$.

 Then it is easy to see that under
the transformation $G^a\rightarrow {\tilde G}^{a}$ the
numerator and denumerator in (4.3.6) are multiplied by the
${\rm Det} {\partial {\tilde G}\over\partial G}$ in the case if
(4.3.5) and (4.3.7) hold.
It is easy to see
(see for details Section 3, eq.(3.2.5)--(3.2.7)) that (4.3.6)
corresponds to (4.3.4) on Lagrangian surfaces if we put
                    $$
F_i=\theta_i-{\partial \Psi (x)\over\partial x^i}
                                   \eqno(4.3.10)
                     $$
(Compare with (4.3.2))

In this case the functional (3.2.3) on Langrangian surface
(4.3.10) is equal to
                      $$
        \Phi _{\tilde {A}}(\Lambda)=
        \int \sqrt{\rho}\prod_a \delta (F^a)
 dx^1 \dots dx^N d\theta_1\dots d\theta_N
                                        \eqno(4.3.11)
                       $$

    (We come immediately from (4.3.6) to (4.3.11) choosing
     $G^i=x^i$ in (4.3.6).)

  To prove that this functional is invariant under infinitesimal
  variation of Lagrangian surface
   $\Lambda\to \Lambda+\delta\Lambda$ in the case if
 $(dv,\{\,,\,\})$ is $SP$ structure  we note
  that under  the infinitesimal transformation
  $\Psi(x)\rightarrow\Psi(x)+\delta\Psi(x)$ in (4.3.10)
                        $$
              \delta\Phi_{\tilde A}=
                       \int
                 {\partial\sqrt \rho\over \partial \theta^i}
                 {\partial \delta\Psi\over\partial x^i}
                  \prod_a  \delta (F_a)
                  dx^1\dots dx^N d\theta^1\dots d\theta^N.
                                                \eqno(4.3.12)
                        $$
 If $(dv,\{\,,\,\})$ is $SP$ structure then from Theorem follows
that
                        $$
             \Delta_0  \sqrt{\rho }=0
             \quad{\rm so}\quad
              \delta \Phi_{\tilde A}=0\,.
                                        \eqno(4.3.13)
                         $$
  \medskip
 {\bf Remark}. In the case if Langrangian surface is
$(n-k.k)$ dimensional surface in the
superspace $ST^*E^{N-k.k}$ (for arbitrary $1\leq k\leq n$)
 one have consider instead (4.3.6) a density.
                     $$
                     A
                     \left(
  z^A , {\partial F^a\over\partial z^A}\,,
             {\partial F^a\over\partial z^*_A}
             \right)=
              {1\over\sqrt{\rho }}
              {\sqrt{{\rm Ber}
    {\partial (G,F)\over\partial (z,z^* )}}}
    \sqrt{{\rm Ber}\{G^{\tilde a}, F^b\}}
                                           \eqno(4.3.14)
                     $$
     where index ${\tilde a}$ have a reversed parity
     to index $a$.
         \medskip
  The density studied above is very essential for our
  considerations but even in the case of
  $SP$ structure it is not $G^{can}_{dv}$--invariant
  density on all surfaces. We present here the example of
  non--trivial $G^{can}_{dv}$--invariant density of a
  second rank.

 Let a $(N-1.N-1)$--dimensional surface in the superspace
 $E^{N.N}$ is defined by the equations
                 $$
      f=0,\quad \varphi=0\quad
   (f\, {\rm is\,even}\, \varphi {\rm\,is\,odd}).
                                                 \eqno(4.3.15)
                 $$
   $E^{N.N}$is provided with $(dv,\{\,,\,\})$ structure.

  One can consider [35]:
                   $$
                 {\tilde A}=
           {1\over\sqrt{\{f,\varphi\}}}
                  \left(
                  \Delta f-
          {\{f,f\}\over 2\{f,\varphi\}}
                \Delta\varphi-
              {\{f,\{f,\varphi\}\}\over\{f,\varphi\}}-
                  {\{f,f\}\over 2\{f,\varphi\}^2}
              \{\varphi,\{f,\varphi\}\}
                       \right)\,.
                                       \eqno(4.3.16)
                        $$
   (4.3.16)  is $G^{can}_{dv}$--invariant
  semidensity---density of weight $\sigma={1\over 2}$
  (A density have weight $\sigma$ if it multiplies
  by the $\sigma$--th power of Ber in (3.2.4)).
      For example if in the point $z_0$ the functions $f$
      and $\varphi$ defining surface by the equations
      (4.3.15) obey to normalization conditions:
                        $$
         \{f,f\}\vert_{z_0}=
         \{f,\{f,\varphi\}\}\vert_{z_0}=0
                          \eqno(4.3.17)
                        $$
         then
                        $$
                   A\vert_{z_0}=
                   {\Delta f\over\sqrt{\{f,\varphi\}}}
                                        \eqno(4.3.18)
                        $$
   and under the transformation $f\rightarrow \lambda f$,
  $\varphi\rightarrow\mu\varphi$ which does not change (4.3.17),
   (4.3.18) multiplies by the
  $({\lambda\over\mu})^{-{1\over 2}}$---the square root of the
  {\it Berezinian} of this transformation.
  ((4.3.16) can be directly computed from (4.3.18) and (4.3.17)).

 One can show that the density (4.3.16) is unique (up to
 multiplication on a constant) in the class
 of the densities of the rank $k\leq 2$ defined on the
 surfaces of the dimension $(p.p)$
 which are invariant under the transformations preserving
 $(dv,\{\,,\,\})$ structure
  (except the volume form itself) [35].
The semidensity (4.3.16) takes odd values.
 It is exotic analogue of Poincare--Cartan
invariant.---${\tilde A}^2=0$ so it cannot be  integrated
 over surfaces.
           \bigskip
\centerline   { 4.4. $SP$--structure and Batalin--Vilkovisky Formalism}
\medskip

In this subsection we again return to considerations
of the section 2 on the basis of odd symplectic geometry.

The space of fields and antifields described in
 a 2-nd section can be naturally provided
with  odd symplectic structures
$(dv,\{ \ ,\ \})$ which in fact are SP
structures.

We recall that $\ep$ is the space of initial fields $\{\ph^A\}$,
$\ep^e_{min}$ is a space of fields
$\{\ph_{min}^A\}=\{\ph^A,c^\al\}$
(the "ghosts" $c^\al$ have the parity opposite to ${\bf R_\al}) $
and $\ep^e$ is a space of fields

   $\{\Phi^A\}=\{\ph^A,c^\al,\lambda_\al,\nu_\al\},
           \qquad (p(c^\al)=p(\nu_\al)=
           p({\bf R}_\alpha)+1=p(\lambda_\al)+1). $
The space of fields-antifields is nothing but a superspace associated
to cotangent
bundle of a corresponding space of fields (see Section 3).
The superspace $ST^*{\cal E}$ have coordinates $\ph^A,\ph_A^*$.
 Analogously
                $$
           \eqalign
                {
& ST^*\ep_{min}^e \, - \,\,
\{\Phi^A_{min},\Phi^*_{Amin}=\ph^A,c^\al,\ph_A^*,c^*_{\al} \}
        \ \ {\rm and} \cr
& ST^*\ep^e \ - \ \{\Phi^A,\Phi^*_A=\ph^A,c^\al,\lambda_\al,
        \nu_\al,\ph_A^*,c^*_{\al},\lambda^{*\al},\nu^{*\al} \}.
                }
                $$
On the space $\ep$ of initial fields $\{\ph^A\}$ one can consider
two volume forms:
                        $$
        dV_0=\prod_A d\ph^A             \eqno(4.4.1)
                        $$
(canonical one) and
                        $$
        dV=e^{S(\ph)}dV_0             \eqno(4.4.2)
                        $$
related with the action $S(\ph)$ of theory.

The canonical form (4.4.1) is naturally prolonged on
$\ep_{min}^e$ and $\ep^e$:
                $$
                \eqalign
                     {
             dV_{0min}^e&=\prod_{A,\al}d\ph^A dc^\al \cr
dV_{0}^e&=\prod_{A,\al}d\ph^Adc^\al d\lambda_\al d\nu_\al.
                     }
                           \eqno(4.4.3)
                           $$
Using the construction of example 4.2.2 one can consider the
lifting of volume forms (4.4.1)-(4.4.3) on the corresponding
spaces of fields-antifields
                        $$
                        \eqalign
                        {
&d\hat{V}_0|_{ST^*\ep}=\prod_A d\ph^Ad\ph^*_A, \cr
&d\hat{V}_0^e|_{ST^*\ep^e_{min}}=\prod_A d\Phi^A_{min}d\Phi^*_{minA}, \cr
&d\hat{V}_0^e|_{ST^*\ep^e}=\prod_A d\Phi^Ad\Phi^*_A
                        }
                                \eqno(4.4.4)
                        $$
and correspondingly:
                        $$
                        \eqalign
                        {
  &d\hat{V}|_{ST^*\ep}=e^{2S(\ph)}d\hat{V}_0|_{ST^*\ep}\cr
       &d\hat{V}^e|_{ST^*\ep^e_{min}}=
       e^{2S(\ph)}d\hat{V}_0|_{ST^*\ep^e_{min}}\cr
       &d\hat{V}^e|_{ST^*\ep^e}=
       e^{2S(\ph)}d\hat{V}_0|_{ST^*\ep^e}
                        }
                        \eqno(4.4.5)
                        $$
  On the space
$ST^*\ep^e $  ($ST^*\ep^e_{min} $) of fields-antifields there is a
third possibility to consider a volume form
			$$
d\hat{V}^m=e^{2{\cal S}(\Phi,\Phi^*)}d\hat{V}_0^e
				\eqno(4.4.6)
			$$
where ${\cal S}(\Phi,\Phi^*)=S(\ph)+
c^\al {R}^A_\al\ph^*_A+\dots$ is master-action obeying to equation
(2.2.7).

The symplectic structure on $ST^*\ep $,  $ST^*\ep^e $,
($ST^*\ep^e_{min} $)  can be naturally defined by the
construction of example (4.1.1)
($x^i\to\Phi^A,\theta_i=x_i^*\to\Phi_A^*$).
 The (2.2.5) is the corresponding
Poisson bracket.

Using a volume forms (4.4.4) -- (4.4.6) and the odd symplectic
structures we come to different structures ($d\hat{V}_0^e,\ \{ \ ,\ \}$),
($d\hat{V}^e,\ \{ \ ,\ \}$),  ($d\hat{V}^m,\ \{ \ ,\ \}$)
on the space of fields-antifields.

The first two structures are $SP$ structures. (See example 4.2.2 and the
statements i), ii) of Theorem.)

{}From the master-equation (2.2.7) and Theorem (statement iii)) follows that
the
structure
                        $$
              (d\hat{V}^m\,,\,\{\,,\,\})
                                         \eqno(4.4.7)
                        $$
constructed via master-action by (4.4.6)  is $SP$--structure too. So using
the Theorem  one can interpret the
 master-- equation
in the following way:

To  find a volume form $d\hat{V}^m$  in $ST^*\ep^e$ such that it obeys to
initial conditions
                        $$
                   d\hat{V}^m=
                      \left(
              e^{c^\al R_\al^A\ph^*_A+\dots}
                     \right)
                      d{\hat V}
                                \eqno(4.4.8)
                        $$
and there are Darboux coordinates (of course non-local) in which
                        $$
                        d\hat{V}^m=1\cdot d\hat{V}_0
                                                \eqno(4.4.9)
                        $$
(Action "dissolves".)
             \medskip
The basic formula (2.2.10) for reduced partition function is
interpreted in a following  way.

To $SP$ structure ($d\hat{V}^m,\ \{ \ ,\ \}$)  on  $ST^*\ep$
corresponds $D$-density (4.3.6).
To this $D$-density corresponds functional
(3.2.3)--- integral over Lagrangian surface
$\Lambda$ in $ST^*\ep$ defined by gauge
conditions (2.2.1), (2.2.9).
                        $$
                    \Psi^\al=0\,
                    \Rightarrow
                    \,\Lambda: \quad
                 \Phi^*_A-
         {\partial (\Psi^\al \nu_\al)\over\partial\Phi^A}=0.
                                        \eqno(4.4.10)
                       $$
The functional (3.2.3) with integrand
 (4.3.6) (with a volume form (4.4.6)) is covariant expression for (2.2.10).
The gauge independence follows from the fact that
($d\hat{V}^m,\ \{ \ ,\ \}$)
structure is $SP$ structure (see (4.3.11)--(4.3.13)).
       \medskip
What is the relation between symmetries of a
theory and $SP$ structure (4.4.7) ?

Let  $\{ R_\al\}$ be basis of symmetries of Theory. (See
subsection 2.1):
                        $$
                        { R}_\al^A{\p S\over\p\ph^A}=0
                                \eqno(4.4.11)
                       $$
(symmetry condition)
                        $$
         (-1)^{p(A)}{\p R_\al^A\over\p\ph^A}=0
                                \eqno(4.4.12)
                        $$
(preserving of canonical volume form (4.4.1)). One can
consider $D$-density
                        $$
                        {\tilde A}=
                 Ber\left({\p\Psi^\al\over\p\ph^A}R^A_\beta\right)
                                \eqno(4.4.13)
                        $$
(See example 3.2.1) and corresponding functional
                        $$
      \Phi=\int Ber\left({\p\Psi^\al\over\p\ph^A}R^A_\beta\right)
        \prod_\al\delta(\Psi^\al)dV.
                                \eqno(4.4.14)
                        $$
In the case if volume form $dV$ in $\ep$ is
 given by (4.4.2) $dV=e^S dV_0$, the
 functional  (4.4.14) is nothing but (2.3.2).

We study the problem of gauge
independence of this functional -- i.e. closeness
of a density (4.4.13).

 Let us consider first

 Toy-example. The number of symmetries is one.
 (4.4.14) is reduced to
                        $$
      \Phi=\int {\p\Psi\over\p\ph^A}R^A \delta(\Psi)e^S
        \prod_A d\ph^A
                        $$
which is nothing but flux of vector field $R$ through a surface
$\Omega:\,\Psi=0$.

 "Gauge" independence means that
                $$
          0= div_{dV}{\bf R}=
         ={\p R^A\over\p\ph^A}+
         {1\over\rho}R^A{\p\rho\over\p\ph^A}=
        {\p R^A\over\p\ph^A}+R^A{\p S\over\p\ph^A},\quad
              (\rho=e^S).
                $$
 which follows from (4.4.11), (4.4.12).

\medskip

In a general case to investigate a problem of
closeness of (4.4.13) we go to BS
representation of this density:
                        $$
                        W_A=\int W_A^e\prod_\al dc^\al
                                        \eqno(4.4.15)
                         $$
where
                        $$
                        W_A^e=e^{c^\al R_\al^A\ph^*_A}
                                        \eqno(4.4.16)
                        $$
is a function on space $ST^*\ep^e_{min}$ (see Example 3.2.1 and
 (2.3.3)).

    \def \ph{\varphi}
    \def\ep{{\cal E}}
 and use the fact that the closeness condition (3.2.13) can be
 expressed in a terms of corresponding $\Delta$ operator.
 (See example 4.2.2).

 For this purpose following to (4.2.7) we consider
 $\Delta$--operator defined on
$ST^*\ep^e_{min} $  by the structure
  $(d{\hat V}, \{\,,\,\})$  where the volume form $d{\hat V}$
 corresponded to $dV$ in (4.4.14) is given by (4.4.5).
 Using that $\{{\bf R}_\alpha\}$ are the symmetries
 (4.4.11) we come to
                       $$
                       \Delta W^e_A=
                       c^\alpha c^\beta
                       \left(
                       t^\gamma_{\alpha\beta}
                       R^A_\gamma
                           +
                    E^{AB}_{\alpha\beta}
                    {\partial S\over\p\ph^B}
                    \right)
                       \ph^*_A
                                          \eqno(4.4.17)
                         $$
           where
                         $$
            [{\bf R}_\alpha,{\bf R}_\beta]=
              t^\gamma_{\alpha\beta}
                       {\bf R}_\gamma +
            E^{AB}_{\alpha\beta}
                    {\partial S\over\partial\ph^B}\,.
                                          \eqno(4.4.18)
                        $$
            (See (2.1.21)).

    In the case if symmetries are abelian
$(t^\gamma_{\alpha\beta}=E^{AB}_{\alpha\beta}=0)$ then
                          $$
                     \Delta W^e_A=0.
                                         \eqno(4.4.19)
                          $$
and evidently in the space $ST^*\ep$
                          $$
                     \Delta W_A=0.
                                         \eqno(4.4.20)
                          $$
 for BS representation (4.4.15) of the density (4.4.13). So this
 density is closed. It means that (4.4.14) is gauge independent.
 (Compare with (2.3.2),(2.3.3)).

 The equation (4.4.19) means that to the function
 $W^e_A$ on $ST^*\ep^e_{min}$ corresponds the closed density
in the space $\ep^e_{min}$ of the fields $\varphi^A$ and ghosts
$c^\alpha$ (if a volume form $dV=e^SdV_0$ in $\ep^e_{min}$).

 $W_A$ is odd function and $W_A^e$ is even one.
  One can see that a volume form
                $$
     d{\hat{\tilde V}}=(W^e)^2 d{\hat V}_0^e=
       e^{2(S(\ph)+c^\alpha R_\alpha^A\ph^*_A)}
                dV_0^e
                        \eqno(4.4.21)
                $$
   provides   $ST^*\ep^e_{min}$    by SP structure because
   $(d{\hat V}, \{\,,\,\})$ is $SP$ structure and $W^e$
    given by (4.4.16) obeys to equation (4.4.19)
    (see the statement iii) of  Theorem).

 We come to $SP$ structure (4.4.6) related with master--action
 in the case where symmetries are abelian (See eq. (2.3.4)).
 \medskip
 To transformation (2.1.23) of the basis
 ${\bf R}_\alpha$ corresponds canonical transformation
                      $$
                      \eqalign
                      {
                      W^e&\rightarrow W^{\prime e}=
                      e^{c^\alpha R_\alpha^A\ph^*_A+\dots}
                      \cr
                      d{\hat v}^e&\rightarrow
           e^{2(S+c^\alpha R_\alpha^A\ph^*_A+\dots)}
                             }
                              \eqno(4.4.22)
                       $$

 In the general case  where initial basis of symmetries is not abelian
 one have to put
   ${\bf {\cal R}}_\alpha$ instead ${\bf R}_\alpha$
   in (4.4.16) where  ${\bf {\cal R}}_\alpha$  is abelian
   (in general non--local) basis of symmetries (2.3.11).
    To the transformation (4.4.22) corresponds
   the transformation from abelian basis to initial one
       performed in a subsection 2.3.
        The function $W^e$ in (4.4.16) plays the role of initial
        conditions. (Compare with (4.4.8)).

 At the end we note that in the case where initial symmetries
 constitute a group (even not--abelian) and
 $\sum_\alpha t^\alpha_{\alpha\beta}=0$ one can see
  by direct computation using (4.4.17)
  that (4.4.20) is obeyed in spite of the
 fact that  (4.4.19) is not obeyed. So the density
 (4.4.13) corresponded to (4.4.15) is closed in this case.

      We come to Faddeev--Popov trick.
      (Compare with (2.3.2) and (2.3.3)).
                        $$ $$
                  \centerline {\bf 5. Acknowledgments}
                         \bigskip
       I am deeply indebted to my collaborator A.P.Nersessian.
       The essential part of the results related to the
       interpretation of BV formalism in terms of Odd
       symplectic geometry is the product of our
       cooperative work.

       I am especially grateful to T.Voronov.
       It would been impossible to write this review without his
       moral support and attention.

   I am grateful to I.Batalin, O.Piguet, I. Tyutin for
fruitful discussions.

 I am grateful  to
 T.S. Akobian, E.V.Tameyan and G.R. Khoudaverdian  who
 helped me  during the preparation of the manuscript of
 this paper.

 And finally I want to use opportunity to express
 my indebtedness to my teacher A.S. Schwarz.

    \vfill
    \eject

\magnification=1200
\def \m {\medskip}
   \centerline {\bf References}

$$ $$

1.Arnold,I.: Mathematical methods of classical mechanics.--
Moscow, Nauka (1974), pp.1--432.
\m
2. Baranov,M.A.,, Schwarz,A.S.--- {\it Funkts. Analiz i ego pril.},
 {\bf 18} No.2,pp.53--54 , (1984).
 \m
3. Baranov,M.A, Schwarz,A.S.: Cohomologies of supermanifolds.
 {\it Funkts. analiz i ego pril.}.  {\bf 18} No.3, pp.69--70,(1984) .
\m
4. Batalin,I.A., Lavrov P., Tyutin I.V---{\it J.Math.Phys.},
   {\bf   31}, 1487. (1990).
   \m
5. Batalin,I.A., Lavrov P., Tyutin I.V---{\it J.Math.Phys.},
   {\bf   32}, 2513. (1991).
   \m
6.  Batalin I.A.,Marnelius.R, Semikhatov,A.---
{\it Preprint of Lebedev Institute}\break FIAN (1995), hep--th 9502031
\m
7.  Batalin,I.A.,Tyutin I.V.:  On possible generalization of
      field-- antifield formalism.
{\it Int. J. of Mod. Phys.} {\bf A8}, No.13, pp.2333--2350 (1993).
\m

8. Batalin,I.A.,Tyutin I.V.:  On the multilevel generalization of
      field-- antifield formalism.
{\it Mod. Phys.Lett.} {\bf A8}, No.38, pp.3673--3681 (1993).
\m
9.   Batalin,I.A.,Tyutin I.V.: On local quantum deformation of
   antisymmetric differential.
   {\it Int. J. of Mod. Phys.} {\bf A9},No.4, pp.517--525,(1994).
   \m
10. Batalin,I.A., Vilkovisky,G.A.:Relativistic $S$--matrix of
dynamical systems
with boson and fermion constraints.
 {\it Phys.Lett.} {\bf 69B}, 309, (1977).
\m
11. Batalin,I.A., Vilkovisky,G.A.: Gauge algebra and quantization.
 {\it Phys.Lett.} {\bf 102B} pp.27--31 (1981).
\m
12. Batalin,I.A., Vilkovisky,G..A.: Closure of the gauge algebra,
 generalized Lie equations and Feynman rules.
 {\it Nucl.Phys.} {\bf B234}, pp.106--124, (1984).
\m
13. Batalin,I.A., Vilkovisky,G.A.: Quantization of Gauge theories
with linearly dependent generators.
 {\it Phys.Rev.} {\bf D28} ,pp.2567--2582, (1983)
 \m
14. Batalin,I.A, Vilkovisky,G.A.: Existence Theorem for gauge
algebra.
 {\it J.Math.Phys.} {\bf 26} pp.172--194 (1985)
 \m
15. Becchi,C., Rouet,A., Stora,R.--- {\it Commun. Math. Phys.}
 {\bf 42},127 (1975)
\m
16. Becchi,C., Rouet,A., Stora,R.: Renormalization of gauge theories.
 {\it Ann. Phys.} (N.Y) {\bf 98}, 287  (1976)
\m
17. Berezin,F.A.: Introduction in analysis with anticommuting
variables.
Moscow Univ., (1983). (English translation is published by Reidel.)
\m
18. Berezin,F.A., Marinov M.S.: Particle spin dynamics as the
  Grassmannian variant of classical mechanics.
{\it Ann. Phys.} {\bf 104}, No.2, pp.336--362 ,(1977).
\m
19. Bernstein,J.N, Leites,D.A.: Integral forms and the Stokes formula
on supermanifolds.
{\it Funkts. Analiz i ego pril.} {\bf 11} No.1 pp. 55--56 (1977)
          \m
20. Bernstein,J.N, Leites,D.A.: How to integrate differential forms on
su\-per\-ma\-ni\-folds.\break
 {\it Funkts.. Analiz i ego pril.} {\bf 11} No.3 pp.70--71
(1977)
\m
21. Browning,A.D., Mc--Millan,D.: The Batalin Fradkin amd Vilkovisky
 Formalism for higher order theories.
  {\it J.Math.Phys.} {\bf 28 }, 438 (1987)
 \m
 22. Buttin,C.---C.R.Acad. Sci. Paris, Ser.A--B,
 {\bf 269} A--87 (1969).
  \m
23.  Dubois--Violette,M.: Systems dynamiques contraints: L'approche
 homologique
 {\it Ann. Inst. Fourier}, {\bf 37},4, pp.45--57, (1987)
\m
24. Fish,J.M.L., Henneaux M.: Homological perturbation theory and the
algebraic structure of the antifield--antibracket formalism for
gauge theories.
 {\it Comm.Math.Phys.} {\bf 128} pp. 627--639 (1990).
 \m
25. Fradkin,E.S, Vilkovisky,G.A.: Quantization of relativistic
systems with constraints.
{\it Phys.Lett.} {\bf 55B}, 224 ,(1975).
\m
26. Fradkin,E., Vilkovisky,G.: Quantization of relativistic systems
with constraints, equivalence of canonical and covariant
formalism in quantum theory of gravitational field.
 {\it CERN preprint} TH-2332 (1977)
\m
27.Fradkin,E.S., Fradkina,T.E.: Quantization of relativistic systems
with boson and fermion first and second class constraints.
{\it Phys. Lett.} {\bf 72B},343, (1978).
\m
28. Gayduk,A.V.,Khudaverdian,O.M, Schwarz,A.S,: Integration on
sur\-faces in \break superspace.
 {\it Theor. Mat. Fiz.} {\bf 52}, pp.375--383, (1982).
\m
29.  Getzler,E.: Batalin--Vilkovisky algebras and two--dimensional
  topological field theory.
   {\it Comm.Math. Phys.} {\bf 159},No.2, 265, (1994).
   \m
30. Hata.,H.,Zwiebach,B.---  {\it Ann. of Phys.},
{\bf 229}, 177, (1994).

    Zwiebach,B.--- preprint IASSNS--HER--92/53.
\m
 31. Henneaux,M.: Hamiltonian form of path integral for a theories
with a gauge freedom.
{\it Phys.Rep.} {\bf 126}, (1985).
\m
32. Henneaux.,M.: Space--time locality of the BRST formalism.
{\it Comm. Math. Phys.} {\bf 140}, pp.1--13 (1991).
\m
 33. Kallosh,R.:  Modified Feynman rules in supergravity.
{\it Nucl. Phys.}, {\bf B141},pp. 141-152 (1978).
\m
 34. Khudaverdian,O.M.: Geometry of superspace provided by Poisson
brackets of different gradings.  {\it J. Math. Phys.}
 {\bf 32}, pp.1934--1937, (1991)
 (Preprint of Geneve Universite. UGVA--DPT 1989/05--613).
\m
 35. Khudaverdian,O.M., Mkrtchian,R.L.: Integral invariants of Buttin
 bracket.
 {\it Lett. Math. Phys.} {\bf 18}, pp. 229--234, (1989).
   ({\it Preprint of Yerevan Physical Institute} ,
   Yerphi, 918(69)--86 (1986),).
\m
 36. Khudaverdian,O.M., Nersessian,A.P.: Formulation of Hamiltonian
mechanics with even and odd Poisson brackets.
{\it Preprint of Yerevan Physical Institute.} Yerphi 1031(81)--87
(1987).
\m
37. Khudaverdian,O.M., Nersessian,A.P.: Poisson brackets of different
grading and strange superalgebras.
{\it J. Math. Phys.}, {\bf 32}, pp.1937--1941, (1991).
 \m
38. Khudaverdian,O., Nersessian,A.  Even and Odd symplectic kahlerian
structures on projective superspace.
{\it J. Math. Phys.} {\bf 34}, pp.5533--5548, (1993).
\m
39. Khudaverdian,O.M., Nersessian,A.P.: On the geometry of
Batalin--Vilkovisky formalism.
 {\it Mod. Phys. Lett.} {\bf A8}, pp. 2377--2385 (1993).
 \m
40. Khudaverdian,O.M, Nersessian,A.P.: Batalin--Vilkovisky formalism
and integration theory on manifolds.
{\it Preprint Geneve Universite} UGVA--DPT 1995/07--896.
VGVA-DPT 1995/07-896
  \m
41. Khudaverdian,O.M, Schwartz, A.S, Tyupkin,Yu.S.:
Integral invariants of supercanonical transformations.
   {\it Lett. Math. Phys.}, {\bf 5} pp.517--520, (1981).
\m
42. Kostant,B., Sternberg,S.: Symplectic reduction, B.R.S.
 cohomology and infinite--di\-men\-si\-onal Clifford algebras.
 {\it Annals of Physics} {\bf 176}, pp.49--113, (1987)
\m
 43. Leites,D.A.: The new Lie superalgebras and Mechanics.
   {\it Docl. Acad. Nauk SSSR} {\bf 236}, pp. 804--807 (1977).
\m
44. Leites,D.A.: Introduction to the theory of supermanifolds.
      {\it Usp. Mat. Nauk} {\bf 35}, No.1, 3--57.
\medskip
45. Leites, D.A.: The theory of Supermanifolds.
   Karelskij Filial AN SSSR (1983).
   \m
 46. Mc--Millan,D.: Yang--Mills theory and the Batalin Fradkin and
 Vilkovisky formalism.
 {\it Journ.Math.Phys.} {\bf 28} , 428 (1987).
\m
 47. Nersessian, A.P: On geometry of supermanifolds with even and
  odd kahlerian structures.  {\it Theor. Math.Phys.}
  {\bf 96}, pp.866--871, (1993).
\m
 48. Nersessian ,A.P.: Antibracket and localization of path integral.
  {\it JETP Lett.} {\bf 58} No.1 pp.66--70. (1993).
  \m
 49.  Nersessian A.P.: From Antibracket to equivariant characteristic
  classes. Preprint JINR, E2--94--377 (1994).
  \m
 50. Nersessian A.P., Damgaard ,P.H.: Comments on the covariant
 $SP(2)$--symmetric lagrangian BRST formalism.
 {\it Phys.Lett.}, {\bf B355}, 150 ,(1995).
\m
 51. Nielsen,N.--- {\it Nucl. Phys.}, {\bf B142}, 306, (1978).
\m
 52. Penkava,M. Schwarz,A.S.: On some algebraical structures arising
 in string theories.  Preprint of Davis University (1992),
  {\it Comm.Math.Phys.}
  \m
 53. Schwarz,A.S.: The partition function of degenerate functional.
 \break {\it Comm..Math.Phys.} {\bf 67}, 1 (1979).
\m
54. Schwarz,A.S.: Supergravity, complex geometry and $G$--structures.
\break
 {\it Comm. Math.Phys.}, {\bf 87}, pp.37--63, (1982)
\m
55. Schwarz,A.S.: Semiclassical approximation in
 Batalin--Vilkovisky formalism. hep--th/9210115
  \m
 56. Schwarz,A.S.; Geometry of Batalin--Vilkovisky Formalism.
{\it Commun. Math. Phys.} {\bf 155},  pp.249--260 (1993).
\m
57. Shander,V.N.: Darboux and Lioville theorems on the supermanifolds.
\break
 {\it Dokl.Akad.Nauk, Bulgary}, {\bf 36}, 309,(1983).
 \m
58. Tyutin,I.V.: Gauge invariance in field theory and statistical
mechanics in the operator formalism.
  {\it Lebedev Physical Institute preprint} FIAN {\bf N39} (1975).
 \m
59. Vandoren, S.,Van Proeyen A.: Simplifications in lagrangian BV
quantization exemplified by the anomalies of chiral
$W_3$ gravity.
 { \it Nucl. Phys.} {\bf B411}  pp.257--306, (1994).
\m
 60. Volkov,D.V., Pashnev A.I.,Soroka,V.A., Tkatch V.I.:
  On a Hamiltonian systems with
 even and odd Poisson brackets and duality of energy
 conservations laws.
 {\it JETP Lett.}, {\bf 44},55, (1986).
 \m
 61. Volkov,D.V.,Soroka V.A.,Tkach,V.I.: On a representation of
 Witten supersymmetric
 quantum mechanics by quantum antibracket.
{\it J.Nucl. Phys.}, {\bf 44}, 810, (1986).
 \m
 62.  Voronov,T.: Theory of integration on vector bundles and
   supermanifolds with application to integral geometry.---
     Baku, International Conference on Topology. Pt.2, p.76 (1987)
\m
63. Voronov,T.: On a class of integral transformations induced
by morphisms of vector bundles.
{\it Mat. Zametki}, {\bf 44},No.6, pp.735--749 (1988).
   \m
 64. Voronov, T.:  Geometric Integration Theory on supermanifolds.
   {\it Sov.Sci.Rev.C Math.} {\bf 9}, pp.1--138  (1992).
      \m
65.  Voronov,T.T., Zorich, A.V.: Complexex of forms on
   supermanifolds.
  {\it Funkts. Analiz i ego pril.}
  {\bf 20}, No.2, pp.58--59., (1988)
   \m
 66. Voronov, T., Zorich ,A.V.:Integral transformations of
 pseudodifferential forms.
 {\it Usp. Mat.Nauk } {\bf 41}, No.6, pp.167--168 (1986).
  \m
 67.  Voronov,T.T., Zorich, A.V.: Cohomologies    of supermanifolds
   and integral geometry.
   {\it  Dokl. Akad. Nauk SSSR.} {\bf 298}, No.3,
  pp.528--533., (1988).
  \m
 68. Voronov,T.T., Zorich, A.V.: Integration on vector bundles.
  {\it Funkts. Analiz i ego pril.} {\bf 22}, No.2, pp.14--25., (1988)
  \m
 69.  de Witt,B.S.: Dynamical theory of groups and fields.
Gordon and Breach. New--York (1965).
\m
70. Witten,E. --- {\it Mod. Phys. Lett.}, {A5}, 487,(1990).

\bye